\documentclass[12pt, fleqn]{article}
\usepackage[cp1251]{inputenc}
\usepackage{latexsym,amsfonts,amssymb}
\usepackage{graphicx}

\usepackage{amsbsy}
\usepackage{amsmath}
\usepackage{epsf}
\usepackage{cite}

\usepackage{color}

\newtheorem{theorem}{Theorem}
\newtheorem{definition}{Definition}
\newtheorem{remark}{Remark}
\newcommand{\bt}{\begin{theo}}
\newcommand{\et}{\end{theo}}
\newcommand{\bd}{\begin{displaymath}}
\newcommand{\ed}{\end{displaymath}}

\newcommand{\be} {\begin{equation}}
\newcommand{\ee} {\end{equation}}
\newcommand{\ba} {\begin{array}}
\newcommand{\ea} {\end{array}}
\newcommand{\bea}{\begin{eqnarray}}
\newcommand{\eea} {\end{eqnarray}}

\begin{document}

\begin{center}
 {\Large \bf A hunter-gatherer--farmer population model:
 \\
 \vspace{0.3cm}
  new
 conditional symmetries  and   exact solutions  \\ \vspace{0.3cm} with biological
  interpretation}
\medskip

{\bf Roman Cherniha \footnote{\small  Corresponding author. E-mail: r.m.cherniha@gmail.com}}
  {\bf and  Vasyl' Davydovych \footnote{\small  E-mail:davydovych@imath.kiev.ua }}
 \\
{\it ~Institute of Mathematics,  National Academy
of Sciences  of Ukraine,\\
 3, Tereshchenkivs'ka Street, Kyiv 01004, Ukraine
}\\
 \end{center}

 \begin{abstract}
New $Q$-conditional (nonclassical) symmetries and  exact solutions
of the  hunter-gatherer--farmer population  model
  proposed by Aoki, Shida and Shigesada
  ({\it Theor Popul Biol 1996;50:1--17})  are constructed. The main method
   used  for the  aforementioned purposes   is an extension of the nonclassical method for system of
   partial differential equations.
 An analysis of properties of the exact solutions obtained and  their biological interpretation  are carried out.
 New results are compared with those derived in  recent studies
 devoted to the same model.
\end{abstract}

conditional symmetry; nonclassical symmetry.

\section{Introduction} \label{sec-1}

 In this work,  we study the model, which was  suggested in  \cite{ao-sh-shige-96} for modeling
 competition between
  farmers and hunter-gatherers that took place thousands of years ago.   Nowadays this model is extensively
   studied by different mathematical techniques
    \cite{aoki-2020,xiao-2021, elia-mimura-2021,fu-mimura-2021}. In particular, a
   detailed archeological background of the model is presented in \cite{aoki-2020}.   The model reflects the recent  DNA studies, which  have shown that early farming spread through most of Europe by the
range expansion of farmers of Anatolian origin took place
simultaneously with  by the conversion to farming of the European
hunter-gatherers, and have confirmed that these hunter-gatherers
continued to coexist with the incoming farmers. It means that three
essentially different populations,  native farmers from Anatolia,
converted farmers with origins in Europe  and hunter-gatherers,
coexisted for many years (see \cite{aoki-2020} and references cited
therein).

  This work is a natural
   continuation of  our recent studies  \cite{ch-dav-2017,ch-dav-2021}, in which Lie and
    $Q$-conditional (nonclassical) symmetries of this model were identified and
    exact solutions were constructed. Moreover a biological interpretation of the solutions
     obtained was provided as well. First of all, we remind the reader that
  after relevant re-scaling (see \cite{ch-dav-2017} for details), the model in the 1D approximation
   takes the form of the  nonlinear reaction-diffusion (RD) system
\begin{equation}\label{1-1}\begin{array}{l}  u_t = d_1 u_{xx}+u(1-u-a_1v),\\  v_t =
d_2 v_{xx}+ a_2v(1-u-a_1v)+uw+a_1vw,\\  w_t = d_3
w_{xx}+a_3w(1-w)-a_4uw-a_5vw,  \end{array}\end{equation} where
$u(t,x), \ v(t,x)$ and $w(t,x)$ are nondimensional densities of
populations of  initial farmers, converted farmers, and
hunter-gatherers, respectively (hereinafter the lower subscripts $t$
and $x$  mean differentiation w.r.t. these variables). System
(\ref{1-1}) is called the hunter-gatherer--farmer (HGF) model and is
the main object of investigation in this paper. We naturally assume
that the diffusivities $d_1, \ d_2$ and $d_3$ are positive
constants. Other parameters are nonnegative constant, moreover
$a_1>0$
 (otherwise  system  (\ref{1-1}) has an autonomous equation and this  means that
the other two populations have no impact on the initial farmer
population) and $a_4>0$ (otherwise the carrying capacity of farmers
is zero \cite{ao-sh-shige-96}). Thus, we consider the HGF system
(\ref{1-1}) with the restrictions \begin{equation}\label{1-3} d_1>0,
d_2>0, d_3>0, a_1>0, a_2\geq0, a_3\geq0, a_4>0,
a_5\geq0.\end{equation}

The main aims of this paper are to derive new nonclassical
symmetries and exact solutions of the HGF system (\ref{1-1}),
 analyse the properties of the solutions obtained  and propose their biological interpretation. The main method
  we are using here is an extension of the nonclassical method for  partial differential equations  (PDEs).
  The latter  was firstly suggested by Bluman and Cole  \cite{1-bl-c} and was further developed in many papers
  (see reviews \cite{1-saccomandi05,oliv-21}
   and monographs \cite{ch-se-pl-book,ch-dav-book} for recent citations). The algorithm for finding
    $Q$-conditional symmetries (following \cite{1-f-s-s}, we  use this terminology  instead of
nonclassical symmetries) of a given PDE is based on the classical
Lie method. However, in contrast to the case of Lie symmetry, the
corresponding
  system of determining equations is {\it nonlinear} and its general solution can be found only
  in exceptional cases. If one deals with a system of PDEs then the problem becomes much
  more complicated. As a result, almost all  works devoted to the construction of   $Q$-conditional symmetries
  were published within the last 10--15 years \cite{ch-2010, arrigo2010,torissi2011,bara-19, ch-dav-math, ch-dav-2021,ch-dav-book}.
To the best of our  knowledge, there are only a few  papers devoted
to nonclassical symmetries of PDE systems  published in the early
2000s \cite{bara-02,ch-se-2003,murata-06}.

Recently  \cite{ch-dav-2021},  $Q$-conditional symmetries and exact
solutions were constructed
 for the HGF system (\ref{1-1}) for the first time. However, a so-called `no-go case' was not examined
 therein. Here we aim to examine  this special case as well  and to identify new $Q$-conditional symmetries.
  Moreover, it is shown that
 these  $Q$-conditional symmetries lead to new exact solutions and some of them possess attractive
 properties,
 reflecting   competition between
  farmers and hunter-gatherers.

The remainder of this paper is organized as follows. In
Section~\ref{sec-2}, we introduce a modification of the notion of
the $Q$-conditional symmetry of the first type, which is needed for
the no-go case, and formulate the main theorem. In
Section~\ref{sec-3}, the symmetry operators obtained in
Section~\ref{sec-2}  are used to construct exact solutions of the
HGF system (\ref{1-1}). Analysis  of the solutions derived    in
order to provide a biological interpretation is carried out  as
well. Finally, we briefly discuss the results obtained and compare
them with those derived in other papers.

\section{ Main theoretical results } \label{sec-2}

First of all, we formulate  the main definition used for deriving
new $Q$-conditional symmetries of the HGF system (\ref{1-1}).
Consider an evolution   system of $m$  second-order equations with
two independent $(t, x)$ and $m$ dependent  $u = (u^1, u^2, \dots,
u^m)$ variables
\begin{equation}
u_t^i=F^i \left(t,x, u, u_x, u_{xx}\right), \ i = 1,2, \dots, m, \
m\geq 2. \label{b3}
\end{equation}
Here $F^i$ are smooth functions of the corresponding variables, the
subscripts $t$ and $x$ denote differentiation w.r.t. these
variables, $u_t^i\equiv \frac{\partial u^i}{\partial t}$,
$u_{x}\equiv \frac{\partial u}{\partial x}= \left(\frac{\partial
u^1}{\partial x},\dots ,\frac{\partial u^m}{\partial x}\right),$ and
$u_{xx}\equiv \frac{\partial^2 u}{\partial x^2}=
\left(\frac{\partial^2 u^1}{\partial x^2},\dots ,\frac{\partial^2
u^m}{\partial x^2}\right)$.

Let us consider the general form of a $Q$-conditional symmetry  of
system~(\ref{b3})\,: \begin{equation}\label{b5} Q = \xi^0 (t, x,
u)\partial_{t} + \xi^1 (t, x, u)\partial_{x} +
 \eta^1(t, x, u)\partial_{u^1}+\ldots+\eta^m(t, x, u)\partial_{u^m},\end{equation} where
 $\xi^i(t,x,u)$ and $\eta^k(t,x,u)$ are smooth  functions to-be-determined  using
the well-known criterion (see, e.g., \cite[Chapter 5]{bl-anco-10}).
For the formulation of the criterion, the notion of the second
prolongation of the operator $Q$ is needed. We remind the reader
that the first  prolongation is
\[ \mbox{\raisebox{-1.7ex}{$\stackrel{\displaystyle Q}{\scriptstyle
1}$}}=Q+\rho^1_t\partial_{u^1_{t}}+\ldots+\rho^m_t\partial_{u^m_{t}}+
\rho^1_x\partial_{u^1_{x}}+\ldots+\rho^m_x\partial_{u^m_{x}},\]
hence the second prolongation can be written as
\[ \mbox{\raisebox{-1.7ex}{$\stackrel{\displaystyle Q}{\scriptstyle
2}$}}=\mbox{\raisebox{-1.7ex}{$\stackrel{\displaystyle
Q}{\scriptstyle
1}$}}+\sigma^1_{tt}\partial_{u^1_{tt}}+\ldots+\sigma^m_{tt}\partial_{u^m_{tt}}
+\sigma^1_{tx}\partial_{u^1_{tx}}+\ldots+\sigma^m_{tx}\partial_{u^m_{tx}}+\sigma^1_{xx}\partial_{u^1_{xx}}+\ldots
+\sigma^m_{xx}\partial_{u^m_{xx}},\] where the coefficients $\rho$
and $\sigma$ with relevant subscripts are expressed via the
functions $\xi^0, \ \xi^1$ and $\eta^k$ and their derivatives by the
well- known formulae (see, e.g., \cite{bl-anco-10,ch-dav-book}).
Actually, the formulae  of prolongations of infinitesimal
 operators were constructed
by Sophus Lie in his classical works in the 1880s
\cite{lie-81,lie-85}.
\begin{definition}\label{bd1}
Operator (\ref{b5})  is called  a $Q$-conditional symmetry
(nonclassical symmetry) for an evolution system of the form
(\ref{b3}) if the invariance criterion
 is  satisfied:
\begin{equation}\label{2-8}  
\mbox{\raisebox{-1.6ex}{$\stackrel{\displaystyle  
Q}{\scriptstyle 2}$}}\, (S_i)\Big\vert_{{\cal{M}}}=0,  \ i = 1,2,
\ldots, m. \end{equation} Here
$\mbox{\raisebox{-1.7ex}{$\stackrel{\displaystyle Q}{\scriptstyle
2}$}}$ is the second prolongation of the operator $Q$ and  the
manifold
  \[{\cal{M}} = \left\{S_i=0,Q\left(u^{i}\right)=0,  \frac{\partial}{\partial t}\,Q\left(u^{i}\right)=0, \frac{\partial}{\partial x}\,Q\left(u^{i}\right)=0,i=1,\dots,m\right\},\] where \[S_i \equiv u_t^i-F^i \left(t,x, u, u_x,u_{xx}\right), \ Q\left(u^{i}\right)\equiv \xi^0{u^i}_t+\xi^1{u^i}_x-\eta^i.\]
\end{definition}

\begin{remark}
In the case of a Lie symmetry operator, the manifold
\[{\cal{M}}_0=\left\{S_i=0,i=1,\dots,m\right\}\supset{\cal{M}} \] should be applied instead
of the Manifold ${\cal{M}}$.
\end{remark}
\begin{remark}
It is shown in \cite[Section 2.3]{ch-dav-book} that the differential
consequences
 $\frac{\partial}{\partial t}\,Q\left(u^{i}\right)=0$ and $\frac{\partial}{\partial x}\,Q\left(u^{i}\right)=0$
  can be ignored when $\xi^0\neq0$ and the system  in question is one of the evolution type.
\end{remark}

It is well-known that solving the problem of constructing
$Q$-conditional symmetries of evolution systems
 depends essentially on the function
 $\xi^0$ because  one should consider two different cases\,:
\begin{enumerate}
  \item $\xi^0\neq0;$
  \item $\xi^0=0, \ \xi^1\neq0.$
\end{enumerate}

Here  we  examine only \textbf{\emph{Case\,2}}, for which the
terminology `no-go case' is often used. Indeed,
\textbf{\emph{Case\,1}} for the HGF system (\ref{1-1}) was
investigated
    in \cite{ch-dav-2021}.
    First of all, we note that the task of constructing  $Q$-conditional symmetries
     with $\xi^0=0$ for scalar evolution equations is reducible  to solving  the equation
     in question \cite{zhdanov}. This statement can be extended on system of evolution equations.
     In other words, it means that application of the invariance criteria (\ref{2-8}) to operator (\ref{b5})
      with $\xi^0=0$ after cumbersome calculations leads to a system of determining equations, which is reducible to (\ref{b3}).
       So, in the case of nonlinear and nonintegrable equations (systems), one can identify only some
        \emph{particular   }   $Q$-conditional symmetries  of the form (\ref{b5})
         with $\xi^0=0$. Notably, even the particular cases obtained by applying Definition~\ref{bd1} may lead to new exact solutions and/or
 can be useful for developing  new techniques such as the algorithm
 of heir equations \cite{nucci-96,hash-nu-13}.

A new algorithm was suggested in \cite{ch-2010}, which allow us to
construct special subsets of $Q$-conditional symmetries in a simpler
way. The algorithm is based on the notion of the $Q$-conditional
symmetry of the $p$-th type ($p=1,\dots,m$). In the case $p=m,$ this
notion leads exactly to the notion of the standard $Q$-conditional
(nonclassical) symmetry. In the case $p=1$, $Q$-conditional
symmetries of the first type are obtained, which form a special
subset of the set of $Q$-conditional  symmetries. It should be
stressed that the no-go case was ignored in \cite{ch-2010}. Recently
\cite{ch-dav-math}, we have shown that the definition of the
$Q$-conditional symmetry proposed in \cite{ch-2010} should be
modified in the no-go case. Having done this, the above mentioned
algorithm allowed us to derive new $Q$-conditional symmetries for
the diffusive Lotka--Volterra (DLV) system. Now we generalize
Definition~2~\cite{ch-dav-math} on any evolution system.
\begin{definition}\label{bd2}
Operator  \begin{equation}\label{b5*} Q = \xi (t, x, u)\partial_{x}
+
 \eta^1(t, x, u)\partial_{u^1}+\ldots+\eta^m(t, x, u)\partial_{u^m}, \ \xi\neq0, \end{equation} is called a $Q$-conditional symmetry
of the first type for an evolution system of the form (\ref{b3}) if
the following invariance criterion
 is  satisfied:
\[  
\mbox{\raisebox{-1.6ex}{$\stackrel{\displaystyle  
Q}{\scriptstyle 2}$}}\, (S_i)\Big\vert_{{\cal{M}}^j_1}=0,  \ i =
1,2, \ldots, m, \] where the Manifold ${\cal{M}}^j_1$ with a fixed
number  $\ j \ ( 1\leq j \leq m)$ is
 \[\left\{S_1=0,S_2=0, \dots, S_m=0, Q\left(u^{j}\right)=0,
 \frac{\partial}{\partial t}\,Q\left(u^{j}\right)=0, \frac{\partial}{\partial x}\,Q\left(u^{j}\right)=0
\right\}.\]
\end{definition}

In the case of evolution system (\ref{b3}), the algorithm of a
complete classification of  $Q$-conditional symmetries of the first
type consists of $m$ steps. The first step reduces to the
application of Definition~\ref{bd2} in the case
\[{\cal{M}}^1_1=\left\{S_1=0, \dots, S_m=0, Q\left(u^{1}\right)=0,
\frac{\partial}{\partial t}\,Q\left(u^{1}\right)=0,
\frac{\partial}{\partial x}\,Q\left(u^{1}\right)=0\right\}\]
 and solving a relevant system of determining equations.
 The next $m-1$ steps are quite similar and one should deal with the manifolds
 ${\cal{M}}^2_1, \dots, {\cal{M}}^m_1$ instead  of ${\cal{M}}^1_1$. If the system in question possesses
  a symmetric structure the number of steps can be reduced. The typical example is
  the two-component DLV system, for which a single step is enough (see \cite{ch-dav-math} for details).
   However, if a given system does not possess a symmetric structure and does not involve
   a subsystem with such structure then the algorithm consists
   of $m$  steps.

Now  we turn back to the HGF system (\ref{1-1}). In this case,
operator (\ref{b5*})  has the form
  \begin{equation} \label{1-2}\begin{array}{l}
 Q=\xi(t,x,u,v,w)\partial_x+\eta^1(t,x,u,v,w)\partial_u\medskip \\ +\eta^2(t,x,u,v,w)\partial_v+\eta^3(t,x,u,v,w)\partial_w.
\end{array}\end{equation}

Our aim is to find all possible $Q$-conditional symmetries of the
first type (\ref{1-2}) for the HGF system (\ref{1-1}).

\begin{remark}In Theorem~\ref{th-1}, the upper indices $u$ and $w$ mean that
the relevant $Q$-conditional symmetry operators satisfy
Definition~\ref{bd2} in the case of the Manifold ${\cal{M}}^1_1$
($u^{1}=u$) and ${\cal{M}}^3_1$  ($u^{3}=w$), respectively.
\end{remark}

\begin{theorem} \label{th-1}
The HGF system (\ref{1-1}) with restrictions (\ref{1-3}) is
invariant under a $Q$-conditional symmetry (symmetries)
 of the first
type (\ref{1-2})
 if and only if the system and the
corresponding symmetry operator(s) have the forms listed below.

\textbf{Case I.} $d_1\neq d_2$\,:
\begin{equation}\label{2-1} \textbf{Subcase I.a} \hskip0.5cm  \begin{array}{l}  u_t = d_1 u_{xx}+u(1-u-a_1v),\\  v_t =
d_2 v_{xx}+ v(1-u-a_1v)+uw+a_1vw,\\  w_t = d_3
w_{xx}+a_3w(1-w)-a_4w(u+a_1v), \ a_3\neq0,
\end{array}\end{equation}
\[Q^u_1=a_1\partial_x+\frac{\alpha_1+2\alpha_2x}{\alpha_0+\alpha_1x+\alpha_2x^2+2d_1\alpha_2t}\,u\left(a_1\partial_u-\partial_v\right);\]

\medskip

\begin{equation}\label{2-5} \textbf{Subcase I.b} \hskip0.5cm \begin{array}{l}  u_t = d_1 u_{xx}+u(1-u-a_1v),\\  v_t =
d_2 v_{xx}+ v(1-u-a_1v)+uw+a_1vw,\\  w_t = d_3 w_{xx}-a_4w(u+a_1v),
\end{array}\end{equation}
\[Q^u_1, \ Q^w_2=\partial_x+e^t(\alpha_0+\alpha_1x)\,w^{\frac{1}{a_4}}\left(a_1\partial_u-\partial_v\right);\]

\medskip

\begin{equation}\label{2-3} \textbf{Subcase I.c} \hskip0.5cm \begin{array}{l}  u_t = d_1 u_{xx}+u(1-u-a_1v), \\  v_t =
d_2 v_{xx}+ a_2v(1-u-a_1v)+uw+a_1vw, \\  w_t = d_3 w_{xx}-uw-a_1vw,
\end{array}\end{equation}
\[Q^u_3=a_1\partial_x+f_1(t,x)\,u\left(a_1\partial_u-\partial_v+(1-a_2)\partial_w\right),\]
\[Q^w_3=(1-a_2)\partial_x+f_3(t,x)\,w\left(a_1\partial_u-\partial_v+(1-a_2)\partial_w\right),\]
where $a_2=\frac{d_2-d_3}{d_1-d_3}$, $\kappa=\frac{1}{\sqrt{\mid
d_1-d_3\mid}}, $ \[f_i(t,x)= \left\{
\begin{array}{l}  \frac{\kappa\left(\alpha_1\cos(\kappa\, x)-\alpha_2\sin(\kappa\, x)\right)}{\alpha_0\exp(d_i\kappa^2t)+\alpha_1\sin(\kappa\, x)+\alpha_2\cos(\kappa\, x)}, \
   \mbox{if} \ d_1>d_3,
 \medskip \\ \frac{\kappa\left(\alpha_1e^{\kappa x}-\alpha_2e^{-\kappa x}\right)}
{\alpha_0\exp(-d_i\kappa^2t)+\alpha_1e^{\kappa x}+\alpha_2e^{-\kappa
x}}, \ \mbox{if} \ d_1<d_3,
\end{array} \right. i=1,3;\]

\medskip \medskip

\textbf{Case II.} $d_1=d_2=1$\,:
\begin{equation}\label{2-2} \textbf{Subcase II.a} \hskip0.5cm \begin{array}{l}  u_t = u_{xx}+u(1-u-a_1v),\\  v_t =
v_{xx}+ v(1-u-a_1v)+uw+a_1vw,\\  w_t = d_3
w_{xx}+a_3w(1-w)-a_4w(u+a_1v), \ a_3\neq0,
\end{array}\end{equation}
\[Q^u_4=\partial_x+g(t,x)\,u\left(a_1\partial_u-\partial_v\right),\]
where $g(t,x)$ is an arbitrary solution of the Burgers equation
$g_t=g_{xx}+2a_1gg_x$;
\begin{equation}\label{2-4} \textbf{Subcase II.b} \hskip0.5cm \begin{array}{l}  u_t = u_{xx}+u(1-u-a_1v),\\  v_t =
v_{xx}+ v(1-u-a_1v)+uw+a_1vw,\\  w_t = d_3 w_{xx}-a_4w(u+a_1v), \
a_4\neq1,  \end{array}\end{equation}
\[Q^u_4, \ Q^w_5=\partial_x+e^t\Big(\alpha_1(1-u-a_1v)+\alpha_2e^{-t}u+\frac{\alpha_1a_1}{1-a_4}w+
h(t,x)w^{\frac{1}{a_4}}\Big)\left(a_1\partial_u-\partial_v\right);\]
\begin{equation}\label{2-6} \textbf{Subcase II.c} \hskip0.5cm \begin{array}{l}  u_t = u_{xx}+u(1-u-a_1v),\\  v_t =
v_{xx}+ v(1-u-a_1v)+uw+a_1vw,\\  w_t = d_3 w_{xx}-uw-a_1vw,
\end{array}\end{equation}
\[Q^u_4, \ Q^w_6=\partial_x+e^t\Big(\alpha_1(1-u-a_1v)+\alpha_2e^{-t}u-\alpha_1a_1w\ln w+h(t,x)w\Big)
\left(a_1\partial_u-\partial_v\right).\] Here $\alpha$ with
subscripts are arbitrary constants, while the function $h(t,x)$ is
an arbitrary  solution of the linear heat equation $h_t=h_{xx}.$
\end{theorem}

\begin{remark} The  functions $f_1$ and $f_3$ reduce to some constants (see the operators $Q^u_3$  and
$Q^w_3$  with $d_1<d_3$)  by setting $\alpha_0=\alpha_1=0$  or
$\alpha_0=\alpha_2=0$.
\end{remark}

\textbf{Sketch of the proof.} In order to derive a complete
classification of  $Q$-conditional symmetries of the first type, we
should  apply the algorithm described above.
 Since system (\ref{1-1}) does not possess  the symmetric structure  the algorithm consist of three steps.
 This means that we should  consider the following three manifolds
\[{\cal{M}}_1^1=\left\{S_1=0,S_2=0,S_3=0, Q(u)=0,  \frac{\partial}{\partial t}\,Q(u)=0, \frac{\partial}{\partial x}\,Q(u)=0
\right\},\]
\[{\cal{M}}_1^2=\left\{S_1=0,S_2=0,S_3=0, Q(v)=0,  \frac{\partial}{\partial t}\,Q(v)=0, \frac{\partial}{\partial x}\,Q(v)=0
\right\}\]  and \[{\cal{M}}_1^3=\left\{S_1=0,S_2=0,S_3=0, Q(w)=0,
\frac{\partial}{\partial t}\,Q(w)=0, \frac{\partial}{\partial
x}\,Q(w)=0 \right\},\] and separately apply Definition~\ref{bd2} for
each manifold. Thus, three different systems of determining
equations should be derived and further solved.

 Let us use the notations \begin{equation}\label{a-26}\begin{array}{l}G^1=u(1-u-a_1v), \\ G^2=a_2v(1-u-a_1v)+uw+a_1vw,
 \\ G^3=a_3w(1-w)-a_4uw-a_5vw\end{array}\end{equation}
 in order to avoid cumbersome formulae.
So,  system (\ref{1-1}) takes the form
\begin{equation}\label{a-1}\begin{array}{l} u_t = d_1
u_{xx}+G^1(u,v,w),\\  v_t = d_2 v_{xx}+G^2(u,v,w),\\  w_t = d_3
w_{xx}+G^3(u,v,w). \end{array}\end{equation}

Applying  Definition~\ref{bd2} with the Manifold ${\cal{M}}_1^1$ to
the RD system (\ref{a-1})  and making straightforward calculations
(see a similar routine in Section 3.3 \cite{ch-se-pl-book}), we
arrive at the following system
 of determining equations\,:
 \begin{eqnarray} && \nonumber \xi_v=\xi_w=\eta^1_{vv}=\eta^1_{ww}=\eta^1_{vw}=\eta^2_{vv}=\eta^2_{ww}=\eta^2_{vw} \medskip \\  && \label{a-2} \hskip2cm=\eta^3_{vv}=\eta^3_{ww}=\eta^3_{vw}=0,  \medskip \\ && \label{a-3}
(d_1-d_2)\eta^1_v=0, \ (d_1-d_3)\eta^1_w=0, \ (d_2-d_3)\eta^2_w=0, \
(d_2-d_3)\eta^3_v=0,
\medskip \\ && \label{a-4}
\xi\eta^1_{xv}+\eta^1\eta^1_{uv}=0, \
\xi\eta^1_{xw}+\eta^1\eta^1_{uw}=0, \
\eta^1\xi_u+\xi\xi_x=0,\medskip \\ && \label{a-5}
(d_1-d_2)\eta^1_w\eta^2_u=2d_2(\xi\eta^2_{xw}+\eta^1\eta^2_{uw}),
(d_1-d_3)\eta^1_v\eta^3_u=2d_3(\xi\eta^3_{xv}+\eta^1\eta^3_{uv}),
\hskip0.5cm
 \medskip \\ && \label{a-6}
2d_2\eta^2_{xv}+G^1\xi_u+\xi_t+\frac{1}{\xi}\Big(d_1\xi_u\eta^1_{x}+
2d_2\eta^1\eta^2_{uv}\Big)+d_1\frac{\eta^1}{\xi^2}\,\xi_u\eta^1_{u}=0,\medskip\\
&& \label{a-7}
2d_3\eta^3_{xw}+G^1\xi_u+\xi_t+\frac{1}{\xi}\Big(d_1\xi_u\eta^1_{x}+2d_3\eta^1\eta^3_{uw}\Big)+
d_1\frac{\eta^1}{\xi^2}\,\xi_u\eta^1_{u}=0,\medskip\\ && \nonumber
d_1\eta^1_{xx}-\eta^1_t-G^1\eta^1_u-G^2\eta^1_v-G^3\eta^1_w+G^1_u\eta^1+G^1_v\eta^2+G^1_w\eta^3
\medskip  \\  && \label{a-8}
+\frac{\eta^1}{\xi}\Big(2d_1\eta^1_{xu}+\xi_t+G^1\xi_u\Big)+
d_1\frac{\eta^1}{\xi^2}\left(\eta^1\eta^1_{uu}+\xi_u\eta^1_x\right)+d_1\frac{{\eta^1}^2}{\xi^3}\,\xi_u\eta^1_{u}=0,\medskip\\
&& \nonumber
d_2\eta^2_{xx}-\eta^2_t-G^1\eta^2_u-G^2\eta^2_v-G^3\eta^2_w+G^2_u\eta^1+G^2_v\eta^2+G^2_w\eta^3
\medskip  \\ && \label{a-9}
+\frac{1}{\xi}\Big(2d_2\eta^1\eta^2_{xu}+(d_2-d_1)\eta^1_x\eta^2_{u}\Big)+
\frac{\eta^1}{\xi^2}\Big(d_2\eta^1\eta^2_{uu}+(d_2-d_1)\eta^1_u\eta^2_{u}\Big)=0,\medskip
\\ && \nonumber
d_3\eta^3_{xx}-\eta^3_t-G^1\eta^3_u-G^2\eta^3_v-G^3\eta^3_w+G^3_u\eta^1+G^3_v\eta^2+G^3_w\eta^3
\medskip  \\  && \label{a-10}
+\frac{1}{\xi}\Big(2d_3\eta^1\eta^3_{xu}+(d_3-d_1)\eta^1_x\eta^3_{u}\Big)+
\frac{\eta^1}{\xi^2}\Big(d_3\eta^1\eta^3_{uu}+(d_3-d_1)\eta^1_u\eta^3_{u}\Big)=0.
\end{eqnarray}

Now we present a detailed analysis  of system
(\ref{a-2})--(\ref{a-10}). First of all, we note that
 equations (\ref{a-3}) lead to five essentially different cases, namely: \medskip \\
\textbf{\emph{(i)}} $\eta^1_v=\eta^1_w=\eta^2_w=\eta^3_v=0$ and all
diffusivities $d_1,
\ d_2$ and $d_3$ are arbitrary constants; \medskip \\
\textbf{\emph{(ii)}} $\eta^1_v\neq0 \Rightarrow d_1=d_2$, $\eta^1_w=\eta^2_w=\eta^3_v=0$ and $d_3$ is an arbitrary constant; \medskip \\
 \textbf{\emph{(iii)}} $\eta^1_w\neq0 \Rightarrow d_1=d_3$, $\eta^1_v=\eta^2_w=\eta^3_v=0$ and $d_2$ is an arbitrary constant; \medskip \\
  \textbf{\emph{(iv)}} $\left(\eta^2_w\right)^2+\left(\eta^3_v\right)^2\neq0 \Rightarrow d_2=d_3$,
  $\eta^1_v=\eta^1_w=0$ and $d_1$ is an arbitrary constant; \medskip \\
   \textbf{\emph{(v)}} $\eta^1_v\eta^1_w=0, \  \left(\eta^1_v\right)^2+\left(\eta^1_w\right)^2\neq0$ and
    $\left(\eta^2_w\right)^2+\left(\eta^3_v\right)^2\neq0 \Rightarrow   d_1=d_2=d_3;$ \medskip \\
    \textbf{\emph{(vi)}} $\eta^1_v\eta^1_w\not=0 \Rightarrow$  $d_1=d_2=d_3.$

Consider  case  \textbf{\emph{(i)}}. Integrating the linear
equations  (\ref{a-2}), we calculate that
 the functions $\xi, \ \eta^1, \ \eta^2$ and $\eta^3$ have the form
\begin{equation}\label{a-11}\begin{array}{l}\xi=\xi(t,x,u), \ \eta^1=r^1(t,x,u), \
\eta^2=r^2(t,x,u)+q^2(t,x,u)\,v, \\
\eta^3=r^3(t,x,u)+q^3(t,x,u)\,w,\end{array}\end{equation} where
$\xi, \ r^i, \ q^2$ and $q^3$ are to-be-determined functions. Taking
into account formulae (\ref{a-11}), we note that equations
(\ref{a-5}) vanish, while those from  (\ref{a-4}) reduce to the
single equation $\eta^1\xi_u+\xi\xi_x=0$.

Now one can substitute (\ref{a-26}) and (\ref{a-11}) into  equations
(\ref{a-6})--(\ref{a-10}).  Since the unknown functions $\xi, \ r^1,
\ r^2, \ r^3,\ q^2$ and $q^3$  do not depend on $v$ and $w$, we can
split the equations obtained w.r.t. these variables  and their
products.
 In particular, equation (\ref{a-6}) takes the form
 \[2d_2q^2_{x}+u(1-u-a_1v)\xi_u+\xi_t+\frac{1}{\xi}\Big(d_1\xi_u r^1_{x}+
2d_2r^1q^2_{u}\Big)+d_1\frac{r^1}{\xi^2}\,\xi_ur^1_{u}=0. \]
 Splitting the last equation w.r.t. the variable $v$, one immediately obtains  $\xi_u=0$ (see restrictions (\ref{1-3})), therefore
 $\xi=\xi(t)$. So, equation (\ref{a-6})
 simplifies to the form
\begin{equation}\label{a-6*} \xi_t+2d_2q^2_{x} +\frac{2d_2}{\xi}r^1q^2_{u}=0.\end{equation}
 Similarly, splitting  equation (\ref{a-9}) w.r.t.
 $vw$ one gets: $q^3=0 \Rightarrow \xi=const$
 (see equation (\ref{a-7})), i.e.
   we can set $\xi=1$ without losing  a generality.  Thus, formulae (\ref{a-11}) take the forms
\begin{equation}\label{a-12}\xi=1, \ \eta^1=r^1(t,x,u), \ \eta^2=r^2(t,x,u)+q^2(t,x,u)\,v, \ \eta^3=r^3(t,x,u).\end{equation}
In other words, the functions (\ref{a-12}) form the general solution
of the subsystem of determining equations consisting
 of  (\ref{a-2})--(\ref{a-7}) with $r^1$ and $q^2$ satisfying (\ref{a-6*}).
  In order to solve the remaining equations (\ref{a-8})--(\ref{a-10}),
  we substitute (\ref{a-26}) and (\ref{a-12}) into
these equations and split the expressions obtained w.r.t. $v$ and
its powers.  As a result, we arrive at \begin{eqnarray} &&
\label{a-13} a_5q^2=0, \ -a_4r^1-a_5r^2-2a_3r^3=0, \
uq^2-r^1-a_1r^2=0,
\medskip \\ && \label{a-14}
ur^1_u-r^1-uq^2=0,   a_1ur^3_u-a_5r^3=0, uq^2_u-a_2q^2=0,
r^1q^2_u+q^2_x=0, \hskip0.4cm
\medskip \\ && \nonumber
d_1r^1_{xx}-r^1_t+d_1\left(r^1\right)^2r^1_{uu}+2d_1r^1r^1_{xu}\medskip \\
&& \label{a-15}
 \hskip1cm+u(u-1)r^1_u+(1-2u)r^1-a_1ur^2=0, \medskip
 \\ && \nonumber
d_2r^2_{xx}-r^2_t+d_2\left(r^1\right)^2r^2_{uu}+2d_2r^1r^2_{xu}+a_2(1-u)r^2+ur^3\medskip \\ && \label{a-16} \hskip1cm+\big(u(u-1)+(d_2-d_1)r^1r^1_u+(d_2-d_1)r^1_x\big)r^2_{u}=0, \medskip \\
&& \nonumber
d_3r^3_{xx}-r^3_t+d_3\left(r^1\right)^2r^3_{uu}+2d_3r^1r^3_{xu}+(a_3-a_4u)r^3\medskip \\ && \label{a-17*} \hskip1cm+\big(u(u-1)+(d_3-d_1)r^1r^1_u+(d_3-d_1)r^1_x\big)r^3_{u}=0,\medskip \\
&& \nonumber
d_2q^2_{xx}-q^2_t+d_2\left(r^1\right)^2q^2_{uu}+2d_2r^1q^2_{xu}+a_1ur^2_u-a_2r^1-2a_1a_2r^2+a_1r^3\medskip
\\ && \label{a-17}
\hskip1cm+\big(u(u-1)+(d_2-d_1)r^1r^1_u+(d_2-d_1)r^1_x\big)q^2_{u}=0.
\end{eqnarray}

Equations (\ref{a-13}) are  algebraic constraints on functions from
(\ref{a-12}). Analyzing  the first equation from (\ref{a-13}), we
need to consider two subcases\,: \textbf{\emph{(i1)}} $a_5\neq0;$
\textbf{\emph{(i2)}} $a_5=0$.

In subcase \textbf{\emph{(i1)}}, we obtain $q^2=0$. Integrating the
first two equations from (\ref{a-14}) and using the last equation
from (\ref{a-13}),  we arrive at the functions $r^1, \ r^2$ and
$r^3$:
\begin{equation}\label{a-18} r^1=f^1(t,x)\,u, \ r^2=-\frac{f^1(t,x)}{a_1}\,u, \ r^3=f^2(t,x)\,u^{\frac{a_5}{a_1}},\end{equation}
where $f^1$ and $f^2$ are to-be-determined smooth functions.

 Substituting $q^2=0$ and (\ref{a-18}) into system (\ref{a-13})--(\ref{a-17}), we
 obtain the system
 \begin{eqnarray} && \label{a-19} (a_1a_4-a_5)f^2=0, \ (a_1a_4-a_5)f^1\,u+2a_1a_3f^2\,u^{\frac{a_5}{a_1}}=0, \medskip \\
 && \label{a-19*}  (a_2-1)f^1\,u+a_1f^2\,u^{\frac{a_5}{a_1}}=0, \medskip \\
 && \label{a-20} f^1_t=d_1f^1_{xx}+2d_1f^1f^1_x,\medskip \\
 && \label{a-21} f^1_t=d_2f^1_{xx}+(3d_2-d_1)f^1f^1_x+(d_2-d_1)\left(f^1\right)^3+(a_2-1)f^1,\medskip \\
 && \nonumber  a_1f^2_t=a_1d_3f^2_{xx}+2a_5d_3f^1f^2_x+a_5(d_3-d_1)f^2f^1_x \medskip \\
 && \label{a-22} +\frac{a_5(a_5d_3-a_1d_1)}{a_1}\left(f^1\right)^2f^2+(a_1a_3-a_5)f^2, \qquad \end{eqnarray}
 which involves three algebraic equations.
Assuming $f^2\neq0$, it is easily shown that equations
(\ref{a-19})--(\ref{a-19*})  produce
\[f^2=\frac{1-a_2}{a_1}\,f^1, \ a_3=0, \ a_4=1, \ a_5=a_1.\]

Now we realize that system (\ref{a-20})--(\ref{a-22})  is an
 overdetermined nonlinear system of PDEs on the function $f^1$. Note that the restriction $d_1\neq d_2$
 should hold  (otherwise  the contradiction
 $(a_2-1)f^1=0$ is obtained, see equations (\ref{a-20})--(\ref{a-21})). It can be  shown by straightforward
 calculations  that
   system (\ref{a-20})--(\ref{a-22}) has nonzero solutions if and only if the restriction
    $a_2=\frac{d_2-d_3}{d_1-d_3}\not=1$ holds.

    Equations (\ref{a-21}) and (\ref{a-22}) coincide under
    the above restriction. Excluding the derivative $f^1_t$ from equation (\ref{a-20}) and substituting
    into (\ref{a-21}) we arrive exactly at  the equation
 \begin{equation}\label{a-23} f^1_{xx}+3f^1f^1_x+\left(f^1\right)^3+\frac{1}{d_1-d_3}\,f^1=0.\end{equation}
 It is well-known (see, e.g., \cite{kamke}) that the nonlinear
equation (\ref{a-23}) is reducible to the linear third-order
ordinary differential equation (ODE)
\begin{equation}\label{a-24} (d_1-d_3)g_{xxx}+g_x=0,\end{equation}
 by the nonlocal substitution $f^1=\frac{g_x}{g}$, where $g(t,x)$ is a new unknown  function.
 Integrating the linear equation (\ref{a-24}),
  we derive two types of its general solutions depending on the sign of $d_1-d_3.$
  Substituting each of them into equation (\ref{a-20}), we obtain two forms of the function $f_1$
   listed in Theorem~\ref{th-1}.
 Thus, we arrive at the operator $Q^u_3$ of the HGF system (\ref{2-3}).

Now we assume that $f^2=0$ and $f^1\neq0$ (for $f^1=0$ the Lie
symmetry operator $\partial_x$ is obtained)  and immediately  arrive
at the restrictions $a_5=a_1a_4, \ a_2=1.$ In this case, system
(\ref{a-20})--(\ref{a-22}) is reducible  to \begin{equation}\label{a-25}\begin{array}{l} f^1_t=d_1f^1_{xx}+2d_1f^1f^1_x,\medskip \\
(d_1-d_2)\left(f^1_{xx}+3f^1f^1_x+\left(f^1\right)^3\right)=0.\end{array}\end{equation}
Now one realizes that the above system has the same structure as
that integrated above. Solving system (\ref{a-25}) and taking into
account (\ref{a-12}) and (\ref{a-18}), we obtain operator $Q^u_1$ of
the HGF system (\ref{2-1}) (in the case $d_1\neq d_2$) and operator
$Q^u_4$ of system (\ref{2-2}) (in the case $d_1=d_2$). Thus, case
$\emph{\textbf{(i)}}$ is completely investigated and the operators
$Q^u_1, \ Q^u_3$ and $Q^u_4$ are constructed.

Cases $\emph{\textbf{(ii)}}$--$\emph{\textbf{(vi)}}$ were also
studied.
 It was proved that     new   $Q$-conditional
symmetry operators are not obtainable.


Applying Definition~\ref{bd2} in the case of the Manifold
${\cal{M}}^3_1$ in a quite similar way,  the operators $Q^w_2, \
Q^w_3, \ Q^w_5$ and $Q^w_6$  were  identified for systems
(\ref{2-5}), (\ref{2-3}), (\ref{2-4}) and (\ref{2-6}), respectively.


 Finally, it was checked by applying
Definition~\ref{bd2} in the case of the Manifold ${\cal{M}}^2_1$
that the HGF system (\ref{1-1}) does not admit new $Q$-conditional
symmetry operators.

{\it The sketch of the proof is now completed.} \hfill $\square$

\begin{remark}
The system of determining equations (\ref{a-2})--(\ref{a-10}) is
valid for any RD system of the form~(\ref{a-1}).
\end{remark}

Now we present the following observation.  All the HGF systems
presented in \textbf{\emph{Case\,I}} of Theorem~\ref{th-1} admit
only a trivial Lie symmetry generated by the operators of time and
space translations (see, Theorem~2.1 \cite{ch-dav-2017}). All the
HGF systems listed in \textbf{\emph{Case\,II}} of Theorem~\ref{th-1}
admit nontrivial Lie symmetries, which can be directly obtained from
the relevant $Q$-conditional symmetries. Indeed, the HGF systems
(\ref{2-2}), (\ref{2-4}) and (\ref{2-6}) admit the Lie symmetry
operator $\partial_x+\alpha u (a_1\partial_u-\partial_v)$ (see Case
4 of Table 1 \cite{ch-dav-2017}) that follows from the operator
$Q^u_4$ if one sets $g(t,x)=\alpha$.  As follows from Case 9 of
Table 1 \cite{ch-dav-2017}, the HGF systems (\ref{2-4}) and
(\ref{2-6}) with
 $d_3=1$ additionally admit the Lie symmetry operators $\partial_x+\alpha
e^t\Big(1-u-a_1v+\frac{a_1}{1-a_4}w\Big)\left(a_1\partial_u-\partial_v\right)$
($a_4\not=1$) and $\partial_x+\alpha
e^tw\left(a_1\partial_u-\partial_v\right)$ ($a_4=1$), respectively.
These Lie symmetry operators can be easily obtained as particular
cases from the operators $Q^w_5$ and $Q^w_6$, respectively. This
observation  is in agreement with the conditional symmetry theory,
which says that Lie symmetries should follow from conditional
symmetries as particular cases.

In conclusion of this section, we present a new result about
conditional symmetries of the DLV systems. It can be checked that
the systems arising in \emph{\textbf{Case II}} of Theorem~\ref{th-1}
are reducible to the DLV systems by the transformation
\begin{equation}\label{2-9} u \rightarrow u, \ u+a_1v \rightarrow v,
\ w\rightarrow w. \end{equation} In fact, applying transformation
(\ref{2-9}) to system (\ref{2-2}) and the operator $Q_4^u$, we
obtain the DLV system
\begin{equation}\label{2-7}\begin{array}{l}  u_t = u_{xx}+u(1-v),\\  v_t =
v_{xx}+ v(1-v+a_1w),\\  w_t = d_3 w_{xx}+w(a_3-a_4v-a_3w),
\end{array}\end{equation} and the operator
\begin{equation}\label{2-10}{Q^u_4}^*=\partial_x+g(t,x)\,u\partial_u,\end{equation}
 where the function $g$ is again an arbitrary solution
of the Burgers equation $g_t=g_{xx}+2gg_x.$ In the case $a_3=0,$ the
DLV system (\ref{2-7}) additionally admits
 the $Q$-conditional symmetry operator
 \begin{equation}\label{2-11}{Q^w_5}^*=\partial_x+e^t\Big(\alpha_1(1-v)+\alpha_2e^{-t}u+\frac{\alpha_1a_1}{1-a_4}w+
h(t,x)w^{\frac{1}{a_4}}\Big)\,\partial_u,\end{equation} if
$a_4\neq1$, and
\begin{equation}\label{2-12}{Q^w_6}^*=\partial_x+e^t\Big(\alpha_1(1-v)+\alpha_2e^{-t}u-\alpha_1a_1w\ln w+h(t,x)w\Big)\,\partial_u,\end{equation} if $a_4=1.$

It should be pointed out that operators (\ref{2-10})--(\ref{2-12})
have different structures from those
 constructed in \cite{che-dav2013}. {\it So, we have derived examples of new
  $Q$-conditional (nonclassical) symmetries of the DLV system (\ref{2-7}).}

\section{ Exact solutions and their interpretation } \label{sec-3}

In this section, our aim is to construct new exact solutions of the
HGF system using the conditional symmetries
 obtained above and to suggest their possible biological interpretations. In what  follows, we restrict
 ourselves to two systems, (\ref{2-3})  and (\ref{2-2}).  The first one was examined   because
the corresponding symmetries have the most complicated structure. In
fact, only operators $Q^u_3$ and $Q^w_3$ involve $\eta^1\neq0, \
\eta^2\neq0$ and $\eta^3\neq0$ (see $\eta^i$ in (\ref{1-2})), while
$\eta^3=0$ in all other conditional symmetries.
The HGF system (\ref{2-2})  was examined because one has identical
structure (up to notations)  to the system investigated recently  in
\cite{elia-mimura-2021} and \cite{fu-mimura-2021}. It should be
stressed that all other
  $Q$-conditional symmetry operators listed in Theorem~\ref{th-1} can
be applied to search for exact solutions in the same way.

\subsection{ The HGF system (\ref{2-3})} \label{subsec-3.1}

Let us construct exact solutions of the HGF system (\ref{2-3}) using
the operators $Q^u_3$ and $Q^w_3$. Firstly we note that one can set
$a_1=1$ without losing a generality because of  the transformation
$u \rightarrow u, \ a_1v \rightarrow v, \ a_1w \rightarrow w$, hence
system (\ref{2-3}) and its operators take  the forms
\begin{equation}\label{2-3*}\begin{array}{l}  u_t = d_1 u_{xx}+u(1-u-v), \\  v_t =
d_2 v_{xx}+ \frac{d_2-d_3}{d_1-d_3}\,v(1-u-v)+uw+vw, \\  w_t = d_3
w_{xx}-uw-vw, \end{array}\end{equation}
\[ \begin{array}{l}Q^u_3=\partial_x+f_1(t,x)\,u\left(\partial_u-\partial_v+(1-a_2)\partial_w\right), \\ \medskip
Q^w_3=(1-a_2)\partial_x+f_3(t,x)\,w\left(\partial_u-\partial_v+(1-a_2)\partial_w\right),
\ a_2=\frac{d_2-d_3}{d_1-d_3}. \end{array}\]

  In order to construct the  ansatz generated by the operator  $Q^u_3$,  according to the standard procedure
  one needs to use a so-called invariance surface condition. In this case,
  it is the first-order PDE system
\begin{equation}\label{3-1}\begin{array}{l}
u_x=f_1(t,x)\,u, \ v_x=-f_1(t,x)\,u, \
w_x=\frac{d_1-d_2}{d_1-d_3}f_1(t,x)\,u,\end{array}\end{equation}
where $f_1$ is defined in Theorem~\ref{th-1}.

Depending on the form of the function $f_1$, the integration of
system (\ref{3-1}) leads to  the  ansatz
\begin{equation}\label{3-2}\begin{array}{l}
u(t,x)=\varphi_1(t)\Big(\alpha_0+\alpha_1\sin(\kappa\, x)\exp(-d_1\kappa^2t)+\alpha_2\cos(\kappa\, x)\exp(-d_1\kappa^2t)\Big),\medskip \\ v(t,x)=\varphi_2(t)-u(t,x),\medskip \\
 w(t,x)=\varphi_3(t)+\frac{d_1-d_2}{d_1-d_3}\,u(t,x),\end{array}\end{equation}
if $d_1>d_3$, and the ansatz
\begin{equation}\label{3-3}\begin{array}{l}
u(t,x)=\varphi_1(t)\Big(\alpha_0+\alpha_1\exp(\kappa\,x+d_1\kappa^2t)+\alpha_2\exp(-\kappa\,x+d_1\kappa^2t)\Big),\medskip
\\ v(t,x)=\varphi_2(t)-
u(t,x),\medskip \\
 w(t,x)=\varphi_3(t)+\frac{d_1-d_2}{d_1-d_3}\,u(t,x),\end{array}\end{equation}
if $d_1<d_3$. Here $\varphi_1(t), \ \varphi_2(t)$ and $\varphi_3(t)$
are new unknown functions, while  $\kappa=\frac{1}{\sqrt{\mid
d_1-d_3\mid}}.$

Substituting  ansatz (\ref{3-2}) into the HGF system (\ref{2-3*}),
we arrive at the ODE system
\begin{equation}\label{3-4}\begin{array}{l}
\varphi_1'+\varphi_1(\varphi_2-1)=0,\medskip\\
\varphi_2'-\varphi_2\varphi_3+
\frac{d_2-d_3}{d_1-d_3}\,\varphi_2(\varphi_2-1)-\frac{\alpha_0(d_1-d_2)}{d_1-d_3}\,\varphi_1=0,
\medskip\\
\varphi_3'+\varphi_2\varphi_3+\frac{\alpha_0(d_1-d_2)}{d_1-d_3}\,\varphi_1=0.
\end{array}\end{equation}
It turns out that ansatz (\ref{3-3}) leads to the same ODE system.

The ODE system  (\ref{3-4}) is nonlinear and its complete
integration is beyond the scope of this work. However, we were able
to construct particular solutions of (\ref{3-4}). It turns out that
the solutions obtained lead to those of the HGF system (\ref{2-3*}),
which possess highly attractive properties.

First of all, we note that the ODE system (\ref{3-4}) possesses two
steady-state points
\[ (\varphi_1,\varphi_2,\varphi_3)= (0,0,w_0),
 \quad (\varphi_1,\varphi_2,\varphi_3)= \left( u_0, 1, \alpha_0 u_0\frac{d_2-d_1}{d_1-d_3} \right), \]
where $u_0$  and $w_0$ are arbitrary parameters. The first
steady-state point leads to a trivial solution of the  the HGF
system (\ref{2-3*}), however, the second one, after substituting
into (\ref{3-2}) and (\ref{3-3}), produces new four-parameter
families of exact solutions. Ansatz (\ref{3-2})  produces the
solutions of the form \begin{equation}\label{3-2*}\begin{array}{l}
u(t,x)=u_0\Big(\alpha_0+\alpha_1\sin(\kappa\, x)\exp(-d_1\kappa^2t)+\alpha_2\cos(\kappa\, x)\exp(-d_1\kappa^2t)\Big),\medskip \\
v(t,x)= 1-u_0\Big(\alpha_0+\alpha_1\sin(\kappa\, x)\exp(-d_1\kappa^2t)+\alpha_2\cos(\kappa\, x)\exp(-d_1\kappa^2t)\Big),\medskip \\
 w(t,x)=u_0\frac{d_1-d_2}{d_1-d_3}\Big(\alpha_1\sin(\kappa\, x)\exp(-d_1\kappa^2t)+
 \alpha_2\cos(\kappa\, x)\exp(-d_1\kappa^2t)\Big),\end{array}\end{equation}
 while  ansatz (\ref{3-3}) leads to those of the form
\[\begin{array}{l}
u(t,x)=u_0\Big(\alpha_0+\alpha_1\exp(\kappa\,x+d_1\kappa^2t)+\alpha_2\exp(-\kappa\,x+d_1\kappa^2t)\Big),\medskip \\
v(t,x)=1-u_0\Big(\alpha_0+\alpha_1\exp(\kappa\,x+d_1\kappa^2t)+\alpha_2\exp(-\kappa\,x+d_1\kappa^2t)\Big),\medskip \\
 w(t,x)=u_0\frac{d_1-d_2}{d_1-d_3}\Big(\alpha_1\exp(\kappa\,x+d_1\kappa^2t)+\alpha_2\exp(-\kappa\,x+d_1\kappa^2t)\Big),\end{array}\]
 where $\alpha_i$  and $u_0$ are arbitrary parameters.

 For example, let us consider a particular case  $ \alpha_2=0$ 
  and assume that  three
populations of initial farmers, converted farmers, and
hunter-gatherers  are interacting in the domain
\[\Omega_{\kappa}=\left\{ (t,x) \in (0,+ \infty)\times \Big(\frac{2k\pi}{\kappa},\frac{(2k+1)\pi}{\kappa}\Big)
\right\}, \  k \in \mathbb{Z}.\] Now we observe  that the exact
solution  (\ref{3-2*})  takes the form
\begin{equation}\label{3-2**}\begin{array}{l}
u(t,x)=u_0\Big(\alpha_0+\alpha_1\sin(\kappa\, x)\exp(-d_1\kappa^2t))\Big),\medskip \\
v(t,x)= 1-u_0\alpha_0-u_0\alpha_1\sin(\kappa\,
x)\exp(-d_1\kappa^2t),
\medskip \\
 w(t,x)=u_0\alpha_1\frac{d_1-d_2}{d_1-d_3}\sin(\kappa\, x)\exp(-d_1\kappa^2t),\end{array}\end{equation}
 and is nonnegative (the population densities cannot  be negative) in $\Omega_{\kappa}$ provided
 \[ u_0(\alpha_0+\alpha_1)\geq 0, \  1\geq u_0(\alpha_0+ \alpha_1), \ 1\geq u_0\alpha_0,  \ u_0\alpha_1(d_1-d_2)\geq 0. \]
 Moreover solution  (\ref{3-2**}) possesses the  asymptotical behavior
\begin{equation}\label{3-11} (u,\,v,\,w)  \rightarrow
\left(u_0\alpha_0,\,  1-u_0\alpha_0,\,0\right) \ \texttt{as} \ t
\rightarrow +\infty.\end{equation}

\begin{figure}[ht!]
    \begin{center}
        \includegraphics[width=5.5cm]{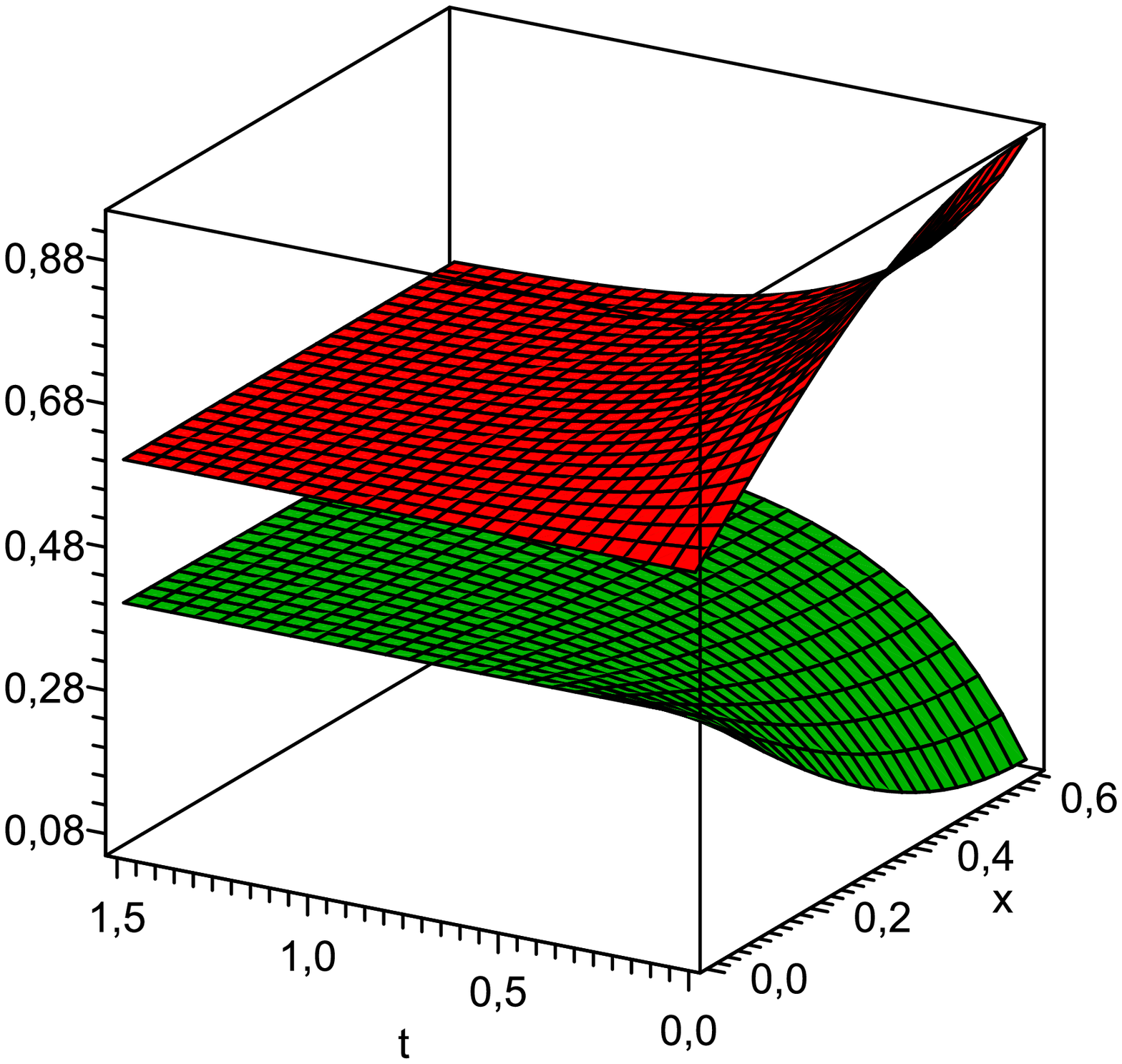}
        \includegraphics[width=5.5cm]{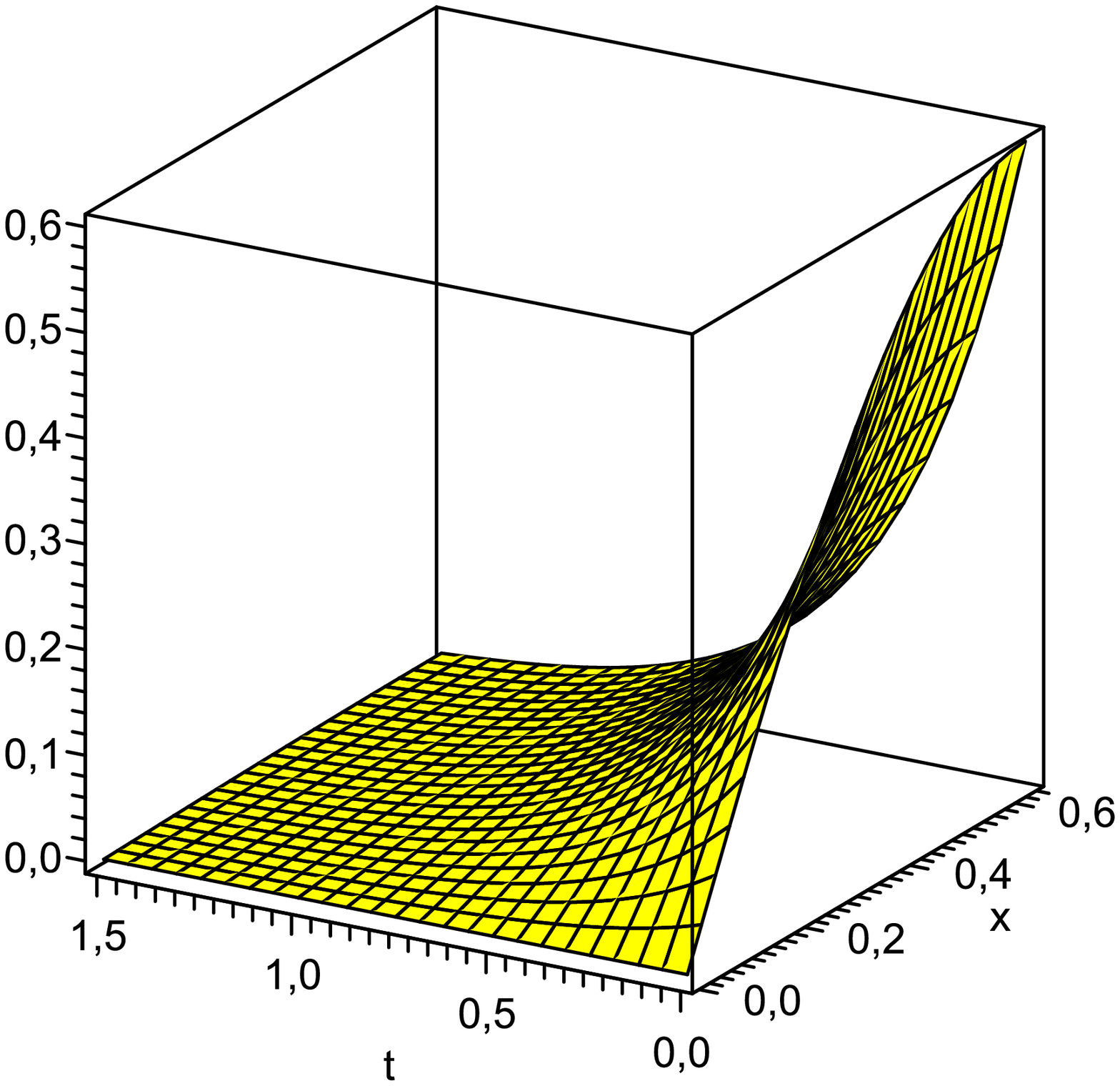}
    \end{center}
    \caption[A]{Surfaces representing the components $u$ (green), $v$ (red) and  $w$ (yellow) of solution (\ref{3-2**}) with $u_0=1, \ \alpha_0=2/5, \ \alpha_1=-1/3, \ \kappa=\sqrt{6}$ of the HGF system  (\ref{2-3*}) with the parameters $d_1=1/2, \ d_2=4/5, \ d_3=1/3$.}
    \label{f1}
\end{figure}

Thus, the exact solution  (\ref{3-2**}) describes such a scenario of
interaction  between  three populations, which leads to the
coexistence of farmers and converted farmers  and to the extinction
of  hunter-gatherers (see an example in Fig.~\ref{f1}). There are
also two special cases. The first one, $u_0\alpha_0=1$, describes
the  scenario leading to a complete extinction of two populations
and only the initial farmers will survive. The second case,
$\alpha_0=0, \  u_0\not=0$, says that eventually all
hunter-gatherers convert into farmers, while all the initial farmers
die out.

Now we present another approach  for  constructing exact solutions
of the ODE system (\ref{3-4}). Let us assume that
$\varphi_2=\beta\varphi_1,$ where $\beta\neq0$ is an arbitrary
constant. In this case, system (\ref{3-4}) can be easily integrated
and has the general  solution
\begin{equation}\label{3-9}\begin{array}{l}
\varphi_1=\frac{Ce^t}{1+\beta C(e^t-1)},\medskip\\
\varphi_2=\frac{\beta Ce^t}{1+\beta C(e^t-1)},
\medskip\\
\varphi_3=\frac{(d_1-d_2)\left(\beta-\alpha_0+\beta
C\left(\alpha_0-\beta-\alpha_0e^t\right)\right)}{\beta(d_1-d_3)\big(1+\beta
C(e^t-1)\big)}.
\end{array}\end{equation}

Taking into account formulae (\ref{3-2}), (\ref{3-9}) and renaming
$C \rightarrow \frac{C}{\beta}, \ \alpha_i \rightarrow
\alpha_i\beta$, the  solution of the HGF system (\ref{2-3*})
\begin{small}
\begin{equation}\label{3-10}\begin{array}{l}
u(t,x)=\frac{Ce^t}{1+C(e^t-1)}\Big(\alpha_0+\alpha_1\sin(\kappa\, x)\exp(-d_1\kappa^2t)+\alpha_2\cos(\kappa\, x)\exp(-d_1\kappa^2t)\Big),\medskip \\ v(t,x)=\frac{Ce^t}{1+C(e^t-1)}\Big(1-\alpha_0-\alpha_1\sin(\kappa\, x)\exp(-d_1\kappa^2t)-\alpha_2\cos(\kappa\, x)\exp(-d_1\kappa^2t)\Big),\medskip \\
 w(t,x)=\frac{d_1-d_2}{(d_1-d_3)\big(1+C(e^t-1)\big)}\Big(1-\alpha_0-C+\alpha_0C+\alpha_1C\sin(\kappa\, x)\exp(-d_3\kappa^2t)+\medskip \\ \alpha_2C\cos(\kappa\, x)\exp(-d_3\kappa^2t)\Big)\end{array}\end{equation} \end{small}
is obtained. Here the coefficients $\alpha_i$  and $C$ are arbitrary
constants, $\kappa=\frac{1}{\sqrt{d_1-d_3}}.$

Assuming the interaction of the populations  in the unbounded domain
 \[\Omega=\left\{ (t,x) \in (0,+ \infty )\times
(-\infty,+ \infty)\right\},\] we note that  the components of
solution (\ref{3-10}) are nonnegative   provided the coefficient
restrictions \[\alpha_0\geq\sqrt{\alpha_1^2+\alpha_2^2},\
1\geq\alpha_0+\sqrt{\alpha_1^2+\alpha_2^2}, \ \left\{
\begin{array}{l}
d_1>d_2, \quad 0\leq C\leq\frac{1-\alpha_0}{1-\alpha_0+\sqrt{\alpha_1^2+\alpha_2^2}},\\
d_1<d_2, \quad
C\geq\frac{1-\alpha_0}{1-\alpha_0-\sqrt{\alpha_1^2+\alpha_2^2}}
\end{array} \right.
\]  hold.
Moreover, the exact solution (\ref{3-10}) possesses the asymptotical
behavior
 \begin{equation}\label{3-11*} (u,\,v,\,w)  \rightarrow
\left(\alpha_0,\,1-\alpha_0,\,0\right) \ \texttt{as} \ t
\rightarrow +\infty.\end{equation}

 Thus, the exact solution (\ref{3-10}) with $\alpha_0$ describes
the same scenario of interaction  of three populations as that does
(\ref{3-2**}), i.e. the coexistence of farmers and converted farmers
and the extinction of  hunter-gatherers. However, in this case, the
interaction can take place both in the   unbounded  domain $\Omega$
and in a bounded domain (w.r.t. the space variable $x$).

\begin{figure}[ht!]
    \begin{center}
        \includegraphics[width=5.5cm]{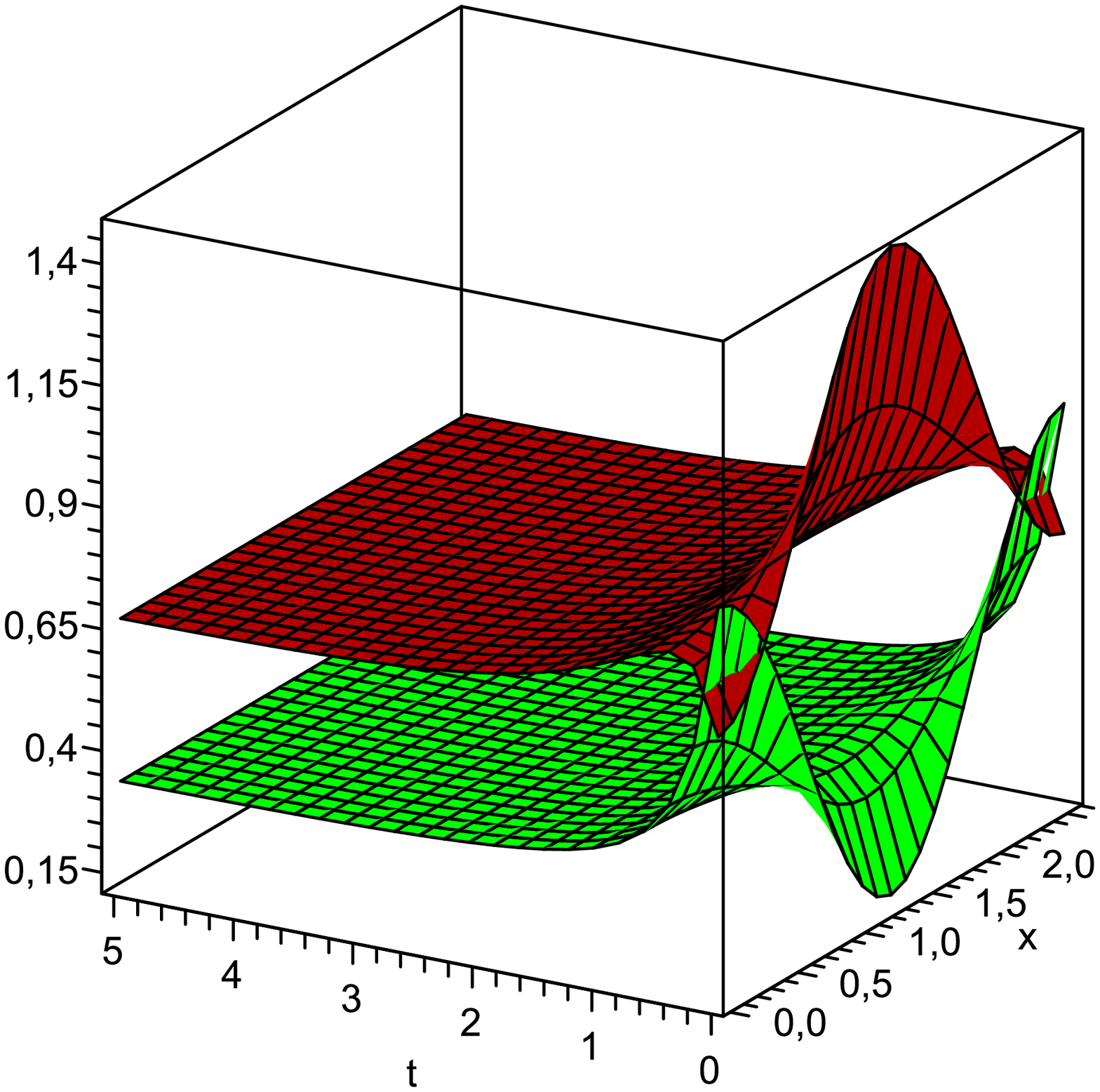}
        \includegraphics[width=5.5cm]{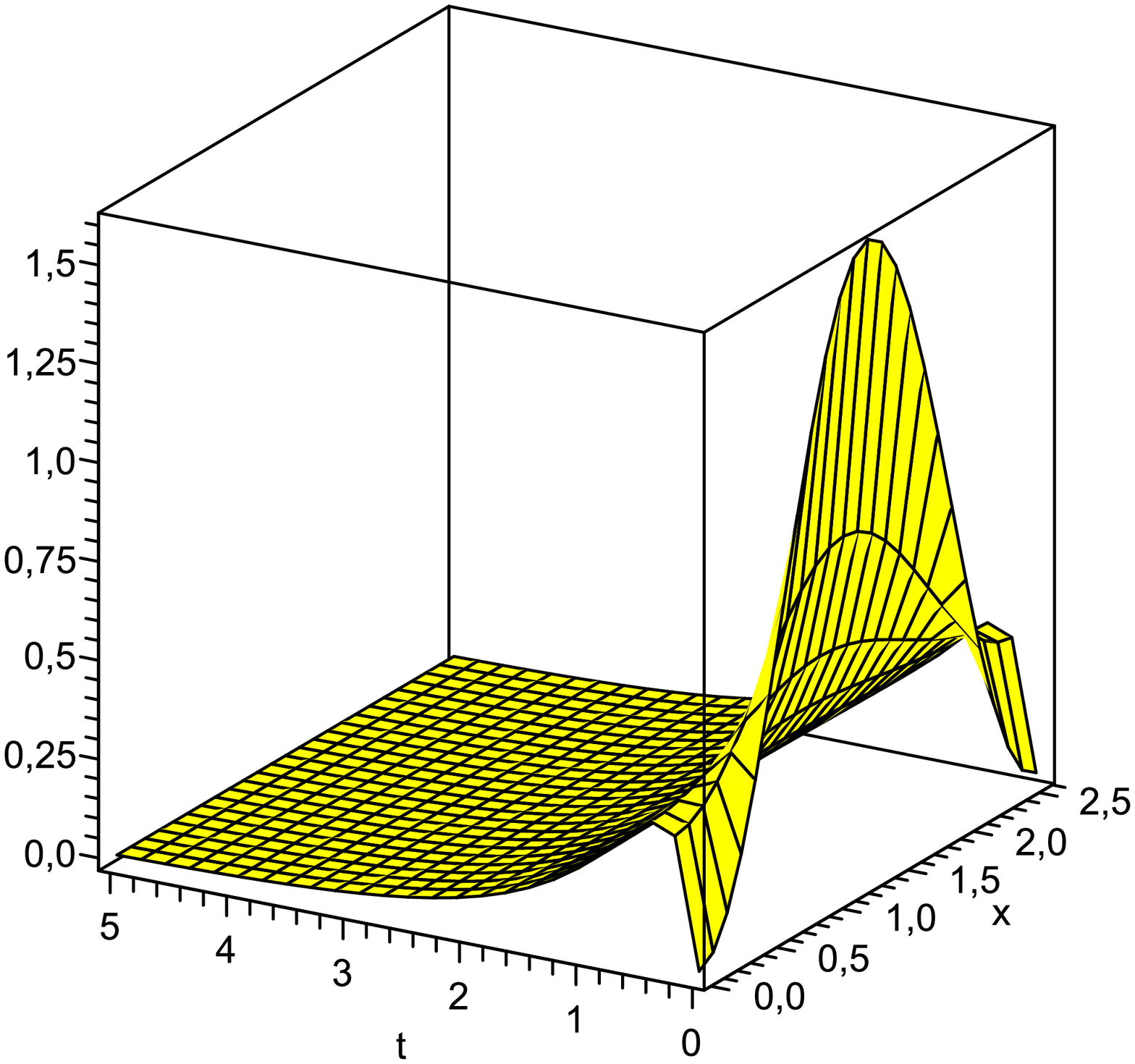}
    \end{center}
    \caption[A]{Surfaces representing the components $u$ (green), $v$ (red) and  $w$ (yellow) of solution  (\ref{3-10}) with $\alpha_0=1/3, \ \alpha_1=0, \ \alpha_2=1/4, \ C=8/5, \ \kappa=\sqrt{6}$ of the HGF system  (\ref{2-3*}) with the parameters $d_1=2/3, \ d_2=1, \ d_3=1/2$.}
    \label{f2}
\end{figure}

Interestingly, the exact solution (\ref{3-10}) with
correctly-specified parameters can be used for solving
boundary-value problems with  typical boundary conditions occurring
in biological problems. Let us consider an example. Assuming that
the population interaction occurs in the bounded domain
\[\Omega_{\kappa}=\left\{ (t,x) \in (0,+ \infty)\times \Big(\frac{2k\pi}{\kappa},
\frac{2(k+1)\pi}{\kappa}\Big) \right\}, \  k \in \mathbb{Z} \] with
no-flux conditions (the zero Neumann conditions)  on the boundaries
\[ \begin{array}{l}
x=\frac{2k\pi}{\kappa}: \ u_x=v_x=w_x=0;\\
x=\frac{2(k+1)\pi}{\kappa}: \ u_x=v_x=w_x=0, \end{array}\]  it  can
be easily identified that the exact solution  (\ref{3-10}) with
$\alpha_1=0$ satisfies these conditions.
 Moreover, if other parameters satisfy the restrictions
\[ 1\geq\alpha_0+\mid \alpha_2\mid \ \geq  2\mid\alpha_2\mid, \quad d_1<d_2, \quad C\geq\frac{1-\alpha_0}{1-\alpha_0-\mid\alpha_2\mid}, \]
then we obtain the plausible picture of the population interaction
(see Fig.~\ref{f2}).

So, we again observe  the coexistence of farmers and converted
farmers and the extinction of  hunter-gatherers as  a  result of the
humanity evolution.


Now we turn back to the ODE system (\ref{3-4}) and show that its
integration reduces
 to a nonlinear second-order ODE. First of all,   the function $\varphi_2$ can be expressed from
the first  equation of (\ref{3-4}):
\begin{equation}\label{3-5}\varphi_2=1-\frac{\varphi_1'}{\varphi_1}.\end{equation}
Substituting (\ref{3-5}) into the third equation of system
(\ref{3-4}) and integrating the equation obtained, we arrive at
\begin{equation}\label{3-6}\varphi_3=\left(C e^{-t}-\frac{\alpha_0(d_1-d_2)}{d_1-d_3}\right)\varphi_1.\end{equation}
Now the second equation  of (\ref{3-4}) can be rewritten in the form
\begin{equation}\label{3-7}\begin{array}{l}
\varphi_1''-\frac{d_1+d_2-2d_3}{d_1-d_3}\frac{{\varphi_1'}^2}{\varphi_1}+
\left(\frac{\alpha_0(d_1-d_2)}{d_1-d_3}-C
e^{-t}\right)\varphi_1\varphi_1'
\\+\frac{d_2-d_3}{d_1-d_3}\varphi_1'+C e^{-t}{\varphi_1}^2=0.
\end{array}\end{equation} So, having the general solution of the
second-order ODE (\ref{3-7}), one easily transforms one into the
general solution of the ODE system (\ref{3-4}) using formulae
(\ref{3-5})--(\ref{3-6}).

ODE (\ref{3-7})  is still  a complicated nonlinear equation. To the
best of our knowledge, its general solution is unknown.
So, we applied additional restrictions  in order to construct
solutions of ODE (\ref{3-7}). For instance, the general solution
 of (\ref{3-7}) with $C=0$ can be derived by reducing to a first-order ODE.
 As a result, we obtain
 \begin{equation}\label{3-12} t+t_0=\int\left(\varphi_1-C_1\varphi_1^{\frac{d_1+d_2-2d_3}{d_1-d_3}}-
 \alpha_0\varphi_1^2\right)^{-1}d\varphi_1.\end{equation}
Here $C_1$ and $t_0$ are  arbitrary constant and the latter  can be
removed by the time shift $t+t_0 \rightarrow t.$ The integral in
(\ref{3-12}) is expressed in terms of elementary functions if some
further  restrictions hold. Examples are presented below.

In the case $C_1=0$, the function $\varphi_1(t)$ has the form
\begin{equation}\label{3-13}\varphi_1(t)=\frac{\pm e^t}{1\pm\alpha_0e^t}.\end{equation}

In the case $C_1\neq0$, we may set
   $2d_3=d_1+d_2$, so that the general solution takes the form
  \begin{equation}\label{3-8} \varphi_1(t)=\left\{
\begin{array}{l}
\frac{1}{2\alpha_0}-\frac{\sqrt{4\alpha_0C_1-1}}{2\alpha_0}\tan\frac{\sqrt{4\alpha_0C_1-1}\,t}{2}, \ \mbox{if} \ 4\alpha_0C_1>1, \ \alpha_0\neq0,\medskip\\
\frac{1}{2\alpha_0}+\frac{\sqrt{1-4\alpha_0C_1}}{2\alpha_0}\tanh\frac{\sqrt{1-4\alpha_0C_1}\,t}{2}, \ \mbox{if} \ 4\alpha_0C_1<1, \ \alpha_0\neq0,\medskip\\
\frac{2+t}{2\alpha_0t}, \ \mbox{if} \ 4\alpha_0C_1=1, \ \alpha_0\neq0,\medskip\\
\pm e^t+C_1, \
   \mbox{if} \  \alpha_0=0.
\end{array} \right.
\end{equation}

Thus, substituting (\ref{3-13}) and (\ref{3-8}) into
(\ref{3-5})--(\ref{3-6}), one easily obtains exact solutions of the
ODE system (\ref{3-4}). Having the known functions $\varphi_i(t), \
i=1,2,3,$ we readily construct the exact solutions of the HGF system
(\ref{2-3}) using formulae  (\ref{3-2}) if $d_1>d_3$ and (\ref{3-3})
if $d_1<d_3.$

Let us consider, for example, the case $4\alpha_0C_1<1$ in detail.
Straightforward  calculations lead to
\begin{equation}\label{3-14}\begin{array}{l}
\varphi_1=\frac{1}{2\alpha_0}+\frac{\sqrt{1-4\alpha_0C_1}}{2\alpha_0}\tanh\frac{\sqrt{1-4\alpha_0C_1}\,t}{2},\medskip\\
\varphi_2=1-\frac{1-4\alpha_0C_1}{2\cosh^2\frac{\sqrt{1-4\alpha_0C_1}\,t}{2}\Big(1+
\sqrt{1-4\alpha_0C_1}\tanh\frac{\sqrt{1-4\alpha_0C_1}\,t}{2}\Big)},
\medskip\\
\varphi_3=-1-\sqrt{1-4\alpha_0C_1}\tanh\frac{\sqrt{1-4\alpha_0C_1}\,t}{2}.
\end{array}\end{equation}

Substituting  (\ref{3-14}) into ansatz (\ref{3-2}), we arrive at the
exact solution
\begin{equation}\label{3-15}\begin{array}{l}
u(t,x)=\left(\frac{1}{2\alpha_0}+\frac{\sqrt{1-4\alpha_0C_1}}{2\alpha_0}\tanh\frac{\sqrt{1-4\alpha_0C_1}\,t}{2}\right)
\Big(\alpha_0+\alpha_1\sin(\kappa\, x)\exp(-d_1\kappa^2t)\\
\hskip2cm+\alpha_2\cos(\kappa\, x)\exp(-d_1\kappa^2t)\Big),\medskip
\\
v(t,x)=1-\frac{1-4\alpha_0C_1}{2\cosh^2\frac{\sqrt{1-4\alpha_0C_1}\,t}{2}\Big(1+
\sqrt{1-4\alpha_0C_1}\tanh\frac{\sqrt{1-4\alpha_0C_1}\,t}{2}\Big)}-
u(t,x),\medskip \\
 w(t,x)=\frac{\exp(-d_1\kappa^2t)}{\alpha_0}\Big(1+\sqrt{1-4\alpha_0C_1}\tanh\frac{\sqrt{1-4\alpha_0C_1}\,t}{2}\Big)\\
 \hskip2cm\times\Big(\alpha_1\sin(\kappa\, x)+\alpha_2\cos(\kappa\, x)\Big),\end{array}\end{equation}
of the HGF system  (\ref{2-3*}) with $d_3=\frac{d_1+d_2}{2}$ and
$d_1>d_2$.

Exact solutions of the form (\ref{3-15}) does not satisfy the
natural requirement of nonnegativity at an arbitrary interval.
However, these solutions are nonnegative provided the interval is
correctly-specified. For instance, setting $\alpha_2=0$  and
renaming $\beta_1=4\alpha_0C_1, \
\beta_2=\frac{\alpha_1}{\alpha_0}$, we transform solution
(\ref{3-15})  into \begin{equation}\label{3-18}\begin{array}{l}
u(t,x)=\left(\frac{1}{2}+\frac{\sqrt{1-\beta_1}}{2}\tanh\frac{\sqrt{1-\beta_1}\,t}{2}\right)
\Big(1+\beta_2\sin(\kappa\, x)\exp(-d_1\kappa^2t)\Big),\medskip \\
v(t,x)=1-\frac{1-\beta_1}{2\cosh^2\frac{\sqrt{1-\beta_1}\,t}{2}\Big(1+
\sqrt{1-\beta_1}\tanh\frac{\sqrt{1-\beta_1}\,t}{2}\Big)}-
u(t,x),\medskip \\
 w(t,x)=\beta_2\sin(\kappa\, x)\exp(-d_1\kappa^2t)\Big(1+\sqrt{1-\beta_1}\tanh\frac{\sqrt{1-\beta_1}\,t}{2}\Big),\end{array}\end{equation}
where $\kappa=\frac{\sqrt{2}}{\sqrt{d_1-d_2}}$,  $\beta_1< 1$ and
$\beta_2$ is an arbitrary constant.

The components of solution (\ref{3-18}) are nonnegative in the
domain \[\Omega_{\kappa}=\left\{ (t,x) \in (0,+ \infty)\times
\Big(\frac{2k\pi}{\kappa},\frac{(2k+1)\pi}{\kappa}\Big) \right\}, \
k \in \mathbb{Z},\]  provided the coefficient restrictions $1
>\beta_1>\beta_2>0$ hold.

Obviously, solution (\ref{3-18})  possesses the asymptotical
behavior \begin{equation}\label{3-16} (u,\,v,\,w)  \rightarrow
\left(\frac{1+\sqrt{1-\beta_1}}{2},\,\frac{1-\sqrt{1-\beta_1}}{2},\,0\right)
\ \texttt{as} \ t  \rightarrow +\infty,\end{equation} which again
implies extinction of the   hunter-gatherer population.

Consider the operator $Q^w_3$. The  ansatz corresponding to this
operator has the form
\begin{equation}\label{3-36}\begin{array}{l}
u(t,x)=\varphi_1(t)+\frac{d_1-d_3}{d_1-d_2}\,w(t,x),\medskip \\ v(t,x)=\varphi_2(t)-\frac{d_1-d_3}{d_1-d_2}\,w(t,x),\medskip \\
 w(t,x)=\varphi_3(t)\Big(\alpha_0+\alpha_1\sin(\kappa\, x)\exp(-d_3\kappa^2t)+\alpha_2\cos(\kappa\, x)\exp(-d_3\kappa^2t)\Big),\end{array}\end{equation}
if $d_1>d_3$, and
\begin{equation}\label{3-37}\begin{array}{l}
u(t,x)=\varphi_1(t)+\frac{d_1-d_3}{d_1-d_2}\,w(t,x),\medskip \\ v(t,x)=\varphi_2(t)-\frac{d_1-d_3}{d_1-d_2}\,w(t,x),\medskip \\
 w(t,x)=\varphi_3(t)\Big(\alpha_0+\alpha_1\exp(\kappa\,x+d_3\kappa^2t)+\alpha_2\exp(-\kappa\,x+d_3\kappa^2t)\Big),\end{array}\end{equation}
if $d_1<d_3$. Here $\varphi_1(t), \ \varphi_2(t)$ and $\varphi_3(t)$
are unknown functions, while  $\kappa=\frac{1}{\sqrt{\mid
d_1-d_3\mid}}.$

Ans\"atze (\ref{3-36}) and (\ref{3-37})  lead to the reduced system
\begin{equation}\label{3-38}\begin{array}{l}
\varphi_1'+\varphi_1\left(\varphi_1+\varphi_2-1\right)-\frac{\alpha_0(d_1-d_3)}{d_1-d_2}\,\varphi_3=0,\medskip\\
\varphi_2'+\frac{d_2-d_3}{d_1-d_3}\,\varphi_2\left(\varphi_1+\varphi_2-1\right)+\frac{\alpha_0(d_2-d_3)}{d_1-d_2}\,\varphi_3=0,
\medskip\\
\varphi_3'+\varphi_3\left(\varphi_1+\varphi_2\right)=0.
\end{array}\end{equation}

Similar to the  ODE system  (\ref{3-4}),  system  (\ref{3-38})  is
nonlinear  and its general solution is unknown. However, some
particular solutions can be derived under additional assumptions.
For example, assuming a linear  functional  dependence between the
functions $\varphi_1$ and $\varphi_2$,
 we have
found the following particular solution
 \begin{equation}\label{3-39}\begin{array}{l}
\varphi_1(t)=\frac{d_1-2d_2+d_3}{2\left(d_1-d_2\right)}\tanh\frac{t}{4},\medskip\\
\varphi_2(t)=\frac{1}{2}+\frac{d_2-d_3}{2\left(d_1-d_2\right)}\tanh\frac{t}{4},
\medskip\\
\varphi_3(t)=\frac{d_1-2d_2+d_3}{2\alpha_0\left(d_1-d_2\right)\left(1+e^{\frac{t}{2}}\right)^2}.
\end{array}\end{equation}

Substituting  the functions $\varphi_1(t)$,  $\varphi_2(t)$ and
$\varphi_3(t)$  into ans\"atze (\ref{3-36}) and (\ref{3-37}), one
obtains two families of exact solutions of the HGF system
(\ref{2-3*}). In particular, the  exact solutions generated by
ansatz  (\ref{3-36})  and formulae  (\ref{3-39}) have the asymptotic
behavior of the form (\ref{3-11}). So, these solutions describe such
interaction  between three populations, which leads to the
coexistence of farmers and converted farmers  and to the extinction
of  hunter-gatherers.

\subsection{ The HGF system (\ref{2-2})} \label{subsec-3.2}

Now we  construct exact solutions of the HGF system (\ref{2-2})
using the operator $Q^u_4$. Applying the transformation $a_1v
\rightarrow v$, we can  rewrite system (\ref{2-2}) and $Q^u_4$ in
the form
\begin{equation}\label{3-21}\begin{array}{l}  u_t = u_{xx}+u(1-u-v),\\  v_t =
v_{xx}+ v(1-u-v)+a_1(u+v)w,\\  w_t = d_3
w_{xx}+a_3w(1-w)-a_4(u+v)w,\end{array}\end{equation}
\[Q^u_4=\partial_x+g(t,x)\,u\left(\partial_u-\partial_v\right),\]
where $g(t,x)$ is an arbitrary solution of the Burgers equation
$g_t=g_{xx}+2gg_x$.

Solving the invariance surface condition for the operator $Q^u_4$,
one obtains the  ansatz
\begin{equation}\label{3-22}\begin{array}{l}  u(t,x)=\varphi_1(t)\exp\left(\int g(t,x)\,dx\right),\\
v(t,x)=\varphi_2(t)-u(t,x),\\
w(t,x)=\varphi_3(t).\end{array}\end{equation}


 It turns out that ansatz (\ref{3-22})  can be rewritten in a simpler  form, using the famous Cole--Hopf substitution \cite{hopf-1950,cole-1951}
 $g=\frac{f_x}{f}$, which reduces the Burgers equation to the linear
 diffusion equation
\begin{equation}\label{3-23**}
f_t=f_{xx}. \end{equation} As a result, ansatz (\ref{3-22}) takes
the form
\begin{equation}\label{3-22*}\begin{array}{l}  u(t,x)=\varphi_1(t)f(t,x),\\
v(t,x)=\varphi_2(t)-  \varphi_1(t)f(t,x),\\
w(t,x)=\varphi_3(t),\end{array}\end{equation} where $f(t,x)$ is an
arbitrary solution of the linear diffusion equation (\ref{3-23**}).
The reduced system corresponding to the ansatz (\ref{3-22*}) has the
form
\begin{equation}\label{3-23*}\begin{array}{l}  \varphi_1'=\varphi_1\left(1-\varphi_2\right),\\  \varphi_2'=\varphi_2\left(1-\varphi_2+a_1\varphi_3\right),\\
\varphi_3'=\varphi_3\left(a_3-a_4\varphi_2-a_3\varphi_3\right).\end{array}\end{equation}

Thus, an arbitrary solution of the linear
 diffusion equation (\ref{3-23**}) generates the exact solution of
 the HGF system (\ref{3-21}) provided  $(\varphi_1, \varphi_2,
 \varphi_3)$ is a solution of the ODE system (\ref{3-23*}).

Let us construct examples of solutions of the nonlinear system
(\ref{3-23*}). Note that this systems contains an autonomous
subsystem for $\varphi_2$ and
 $\varphi_3$.  Because it is nothing else but the two-component Lotka--Volterra
 system (without diffusion), which is nonintegrable,
 we apply a technique used by C and D \cite{ch-du-04} in order to construct
 particular solutions. So, assuming
$\varphi_3=\beta_1\varphi_2+\beta_2,$  with  $\beta_1\neq0$ and
$\beta_2$ are arbitrary constants, system (\ref{3-23*})  can be
easily integrated and has nontrivial solutions in two cases. Having
$\varphi_2$ and  $\varphi_3$ and solving the first ODE from
(\ref{3-23*}), we obtain the first solution
\begin{equation}\label{3-24}\begin{array}{l}
\varphi_1=C_2\,e^t\left(1+C_1\left(e^t-1\right)\right)^{-\frac{1+a_1}{1+a_1a_4}},\medskip\\
\varphi_2=\frac{1+a_1}{1+a_1a_4}\frac{C_1e^t}{1+C_1\left(e^t-1\right)},\medskip\\
\varphi_3=\frac{1-a_4}{1+a_1a_4}\frac{C_1e^t}{1+C_1\left(e^t-1\right)},\end{array}\end{equation}
if $a_3=1$, and the second  solution
\begin{equation}\label{3-25}\begin{array}{l}  \varphi_1=
C_2\,e^t\left(1+C_1\left(e^{(1+a_1)t}-1\right)\right)^{-\frac{1}{1+a_1}},\medskip\\
\varphi_2=\frac{C_1e^{(1+a_1)t}}{1+C_1\left(e^{(1+a_1)t}-1\right)},\medskip\\
\varphi_3=\frac{1-C_1}{1+C_1\left(e^{(1+a_1)t}-1\right)},\end{array}\end{equation}
if $a_3=a_4-a_1-1$ ($C_1$ and $C_2$ are arbitrary constants).

Thus, substituting  the functions $\varphi_1(t)$,  $\varphi_2(t)$
and $\varphi_3(t)$ given  by  formulae (\ref{3-24}) and (\ref{3-25})
into ansatz (\ref{3-22*}), one  immediately obtains two families of
exact solutions of the HGF system (\ref{3-21}) involving arbitrary
solutions of  the linear  diffusion equation.

  Let us consider in detail the solutions of the form
\begin{equation}\label{3-30}\begin{array}{l}
 u(t,x)=C_2\,e^t\left(1+C_1\left(e^t-1\right)\right)^{-\frac{1+a_1}{1+a_1a_4}}f(t,x),\\
v(t,x)= \frac{1+a_1}{1+a_1a_4}\frac{C_1e^t}{1+C_1\left(e^t-1\right)}
- C_2\,e^t\left(1+C_1\left(e^t-1\right)\right)^{-\frac{1+a_1}{1+a_1a_4}}f(t,x),\\
w(t,x)=
\frac{1-a_4}{1+a_1a_4}\frac{C_1e^t}{1+C_1\left(e^t-1\right)},
\end{array}\end{equation}
which arise in the case $a_3=1$.

Note that the exact solution  (\ref{3-30}) includes that constructed
in \cite{ch-dav-2017} (see formula (3.12) therein)   as a particular
case.  In fact, if  one sets $f(t,x)=\exp\left(\alpha^2t+\alpha
x\right)$  ($\alpha$ is an arbitrary constant) in (\ref{3-30}) then
the exact solution from \cite{ch-dav-2017} is immediately obtained.

We assume that  the populations interact  in the bounded domain
\[ \Omega_{ab}=\Big\{ (t,x) \in (0,+ \infty)\times(a,\ b) \Big\}, \  a<b \in \mathbb{R} \] and
no-flux conditions (the zero Neumann conditions)
\begin{equation}\label{3-31}
\begin{array}{l}
x=a: \ u_x=v_x=w_x=0;\\
x=b: \ u_x=v_x=w_x=0 \end{array}\end{equation} are imposed.

According to the classical theory of linear diffusion equations,
there exist a   smooth nonnegative bounded  solution, $f_0(t,x)$, of
equation (\ref{3-23**}) that satisfies the  zero Neumann conditions
\[  x=a: \ f_x=0, \quad  x=b: \ f_x=0 \]
and the initial condition
\[ t=0: \  f = F(x), \]
where  $F(x)$ is an arbitrary smooth function such that  $0\leq
F(x)\leq A$, $A \in \mathbb{R_+}$. Moreover, the solution
$f_0(t,x)$  can be constructed in an explicit form using, e.g., the
Fourier method.

Now we realize that the exact solution  (\ref{3-30}) with
$f(t,x)=f_0(t,x)$  satisfies  the no-flux conditions  (\ref{3-31}).
Moreover, all components are bounded and nonnegative provided
 the  constants in  (\ref{3-30}) are
correctly specified. Indeed, the third component $w$ is smooth,
nonnegative and  bounded  provided $C_1>0$  and $a_4\leq 1$, while
the components  $u$ and $v$  possess the same properties if $C_2\geq
0$ and is sufficiently small. Having the afore-cited restrictions,
we observe the following asymptotical behavior of  the exact
solution (\ref{3-30}) \begin{equation}\label{3-32} (u,\,v,\,w)
\rightarrow
\left(0,\,\frac{1+a_1}{1+a_1a_4},\,\frac{1-a_4}{1+a_1a_4}\right) \
\texttt{as} \ t  \rightarrow +\infty\end{equation} if  $a_4< 1$ and
\[ (u,\,v,\,w)  \rightarrow
\left(\frac{C_2A_{\infty}(x)}{C_1},\,1-
\frac{C_2A_{\infty}(x)}{C_1},\, 0\right) \ \texttt{as} \ t
\rightarrow +\infty\] if  $a_4= 1$, provided the condition $\lim
\limits_{t\rightarrow +\infty} f_0(t,x)= A_{\infty}(x)$ takes place.

Now we  present  a biological meaning of the exact solution
(\ref{3-30}) with $f(t,x)=f_0(t,x)$  as follows.  This  solution
with $a_4< 1$ describes such a scenario of interaction between three
populations, which predicts  the coexistence of converted farmers
and hunter-gatherers and  the total extinction of initial farmers.
This scenario differs from that obtained for the solutions of the
HGF system (\ref{2-3*}) (see formulae (\ref{3-11}), (\ref{3-11*})
and (\ref{3-16})). Interestingly, the asymptotic behavior
(\ref{3-32}) is in agreement with the results derived in
\cite{elia-mimura-2021} and shown numerically in
\cite{fu-mimura-2021}. In fact, the exact solution (\ref{3-30}) is
valid for  the HGF system (\ref{3-21}) with the coefficient
restriction  $a_3=1>a_4$. In other words, the self-reproduction rate
 of  hunter-gatherers described by the coefficient
$a_3$ should be sufficiently high in order to survive. It should be
stressed that there are no examples of exact solutions in
\cite{elia-mimura-2021} and \cite{fu-mimura-2021} but only theorems
of existence and numerical solutions.  Finally, we note that the
density of    hunter-gatherers in (\ref{3-30}) does not depend on
the space variable $x$.  This means biologically  that the
hunter-gatherer diffusion  is very high, therefore  they disperse
uniformly in space, i.e., $w_{xx}\approx 0$.

\begin{figure}[ht!]
    \begin{center}
        \includegraphics[width=5.5cm]{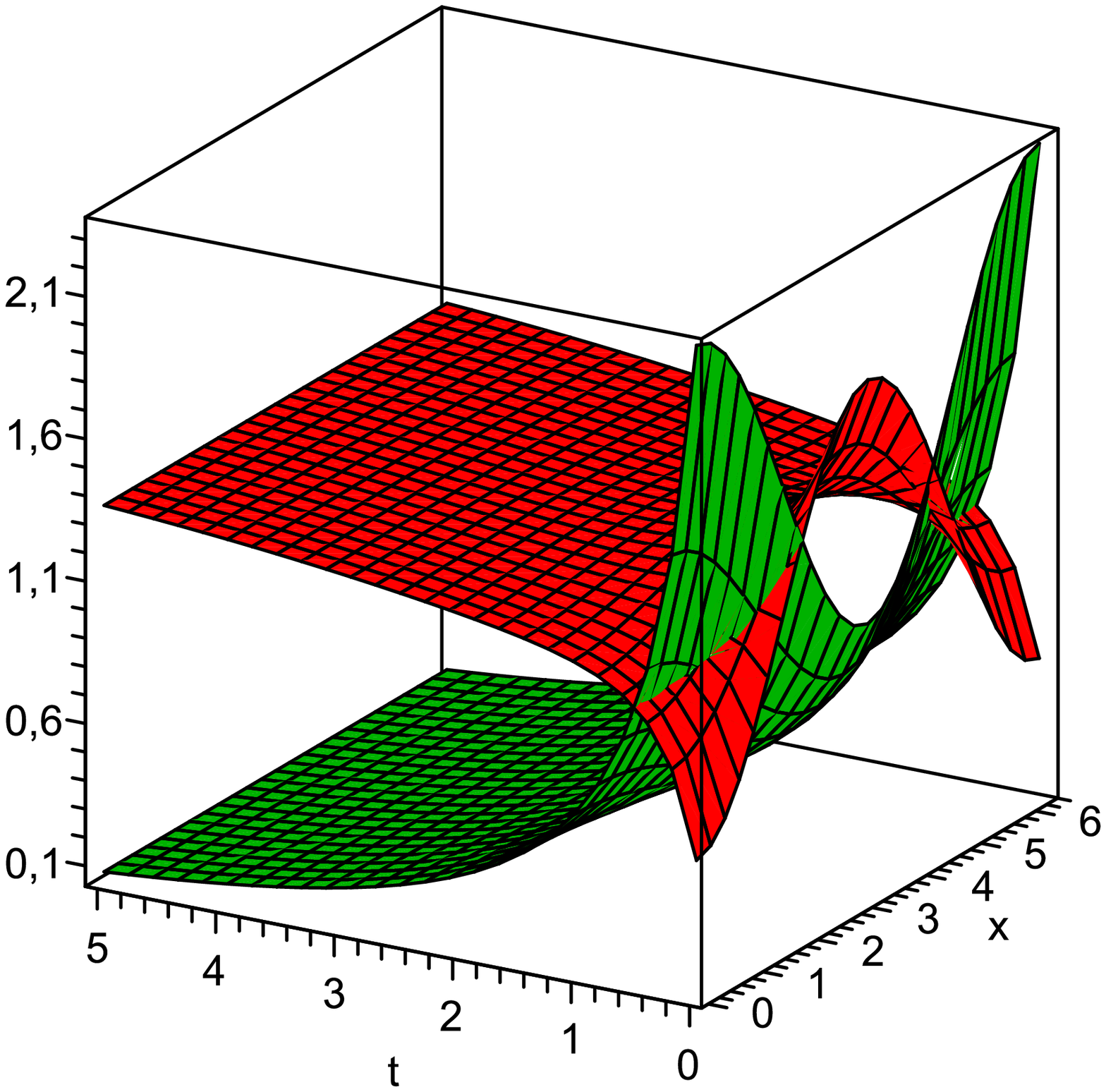}
        \includegraphics[width=5.5cm]{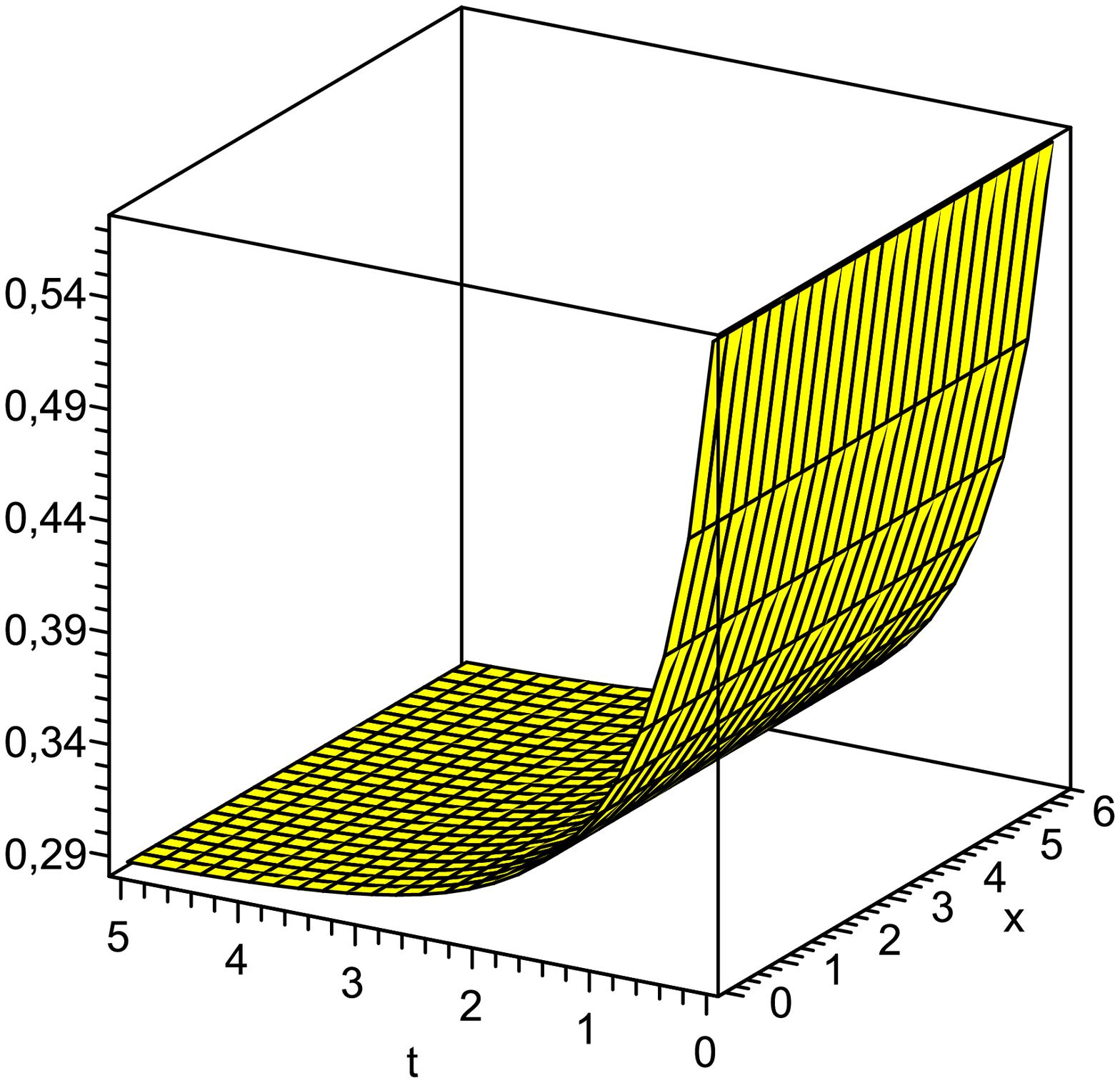}
    \end{center}
    \caption[A]{Surfaces representing the components $u$ (green), $v$ (red) and  $w$ (yellow) of solution  (\ref{3-26}) with $C_1=2, \ C_2=2/3, \ C_3=5/2$ of the HGF system  (\ref{3-21}) with the parameters $a_1=3/2, \ a_3=1, \ a_4=1/2$.}
    \label{f3}
\end{figure}

Setting $F(x)=C_3+\cos x$ (here $C_3\geq1$ is an arbitrary
constant), the function $f_0(t,x)$ can easily be  derived and one
has the form $f_0(t,x)=C_3+e^{-t}\cos x$. In this case, the exact
solution (\ref{3-30}) takes the form
\begin{equation}\label{3-26}\begin{array}{l}  u(t,x)=C_2\left(1+C_1\left(e^t-1\right)\right)^{-\frac{1+a_1}{1+a_1a_4}}\left(C_3e^t+\cos x\right),\medskip\\
v(t,x)=\frac{1+a_1}{1+a_1a_4}\frac{C_1e^t}{1+C_1\left(e^t-1\right)}-C_2\left(1+C_1\left(e^t-1\right)\right)^{-\frac{1+a_1}{1+a_1a_4}}\left(C_3e^t+\cos x\right),\medskip\\
w(t,x)=\frac{1-a_4}{1+a_1a_4}\frac{C_1e^t}{1+C_1\left(e^t-1\right)}.\end{array}\end{equation}
An example of solution (\ref{3-26}) (that is defined in the domain
$\Omega_{2\pi}=\left\{ (t,x) \in (0,+ \infty)\times(0,\ 2\pi)
\right\}$) with correctly-specified coefficients
 is presented in Fig.~\ref{f3}.

Finally, we note that the exact solutions of the form
\begin{equation}\label{3-34}\begin{array}{l}
 u(t,x)=C_2\,e^t\left(1+C_1\left(e^{(1+a_1)t}-1\right)\right)^{-\frac{1}{1+a_1}}f(t,x),\\
v(t,x)=\frac{C_1e^{(1+a_1)t}}{1+C_1\left(e^{(1+a_1)t}-1\right)}
- C_2\,e^t\left(1+C_1\left(e^{(1+a_1)t}-1\right)\right)^{-\frac{1}{1+a_1}}f(t,x),\\
w(t,x)= \frac{1-C_1}{1+C_1\left(e^{(1+a_1)t}-1\right)},
\end{array}\end{equation}
arising  in the case $a_3=a_4-a_1-1$ (see formulae (\ref{3-25})),
can be examined in the same way. As a result,  the   exact solution
(\ref{3-34}) with a correctly-specified function  $f(t,x)=f_0(t,x)$
 (see
the
 previous page about the function $f_0$) has the asymptotical behavior
 \begin{equation}\label{3-35} (u,\,v,\,w) \rightarrow
\Big(C_1^{-\frac{1}{1+a_1}}C_2\
f_0^\infty(x),\,1-C_1^{-\frac{1}{1+a_1}}C_2\ f_0^\infty(x),\,0\Big)
\ \texttt{as} \ t \rightarrow +\infty,\end{equation} where
$f_0^\infty(x)=\lim \limits_{t\rightarrow +\infty} f_0(t,x)$.

Thus, the exact solution (\ref{3-34}) with $f(t,x)=f_0(t,x)$
describes the  scenario of interaction between three populations,
which predicts the coexistence of converted farmers and initial
farmers and  the total extinction of hunter-gatherers.  Thus, it is
the same scenario as one  obtained for  solutions of the HGF system
(\ref{2-3*}).  Moreover, it is in agreement with the theoretical and
numerical results derived in \cite{elia-mimura-2021, fu-mimura-2021}
because $a_3<a_4$.

 In conclusion of this section, we present the
following observation. Here we were looking for  exact solutions of
the HGF systems satisfying no-flux conditions on boundaries in
domains of the form $\Omega_k$ and $\Omega_{ab}$.
 In other words, the corresponding nonlinear   boundary value problems (BVPs) were solved.
 It means, that  the HGF systems  (\ref{2-3*}) and (\ref{3-21}) supplied by  the zero
  Neumann conditions are conditionally invariant w.r.t.  the corresponding $Q$-conditional symmetries. We note that the rigorous  definition of
conditional symmetry of BVP was firstly formulated in
\cite{ch-king-15} and one is a  nontrivial extension of  earlier
definitions of Lie symmetry of BVP \cite{bl-anco02}. However, a
complete description of Lie and conditional symmetries of BVPs with
governing system  (\ref{1-1}) is a highly nontrivial problem and
lies beyond the scope of  paper.

\section{ Discussion } \label{sec-4}

In this work, $Q$-conditional (nonclassical) symmetries of the HGF
system (\ref{1-1}) are constructed in a so-called no-go case. We
point out that the $Q$-conditional symmetries in the regular case,
when $\xi^0\not=0$ in  (\ref{b5}), were earlier identified   in
\cite{ch-dav-2021}. As the no-go case is more complicated, a new
definition was established in order to
 make essential progress in search for
$Q$-conditional  symmetries. Applying a new algorithm based on
Definition~\ref{bd2}, we have proved Theorem~\ref{th-1} giving a
complete description of $Q$-conditional symmetries of the first
type. The symmetries obtained  do not coincide  with those derived
in \cite{ch-dav-2021}, i.e. they are new.

All the  $Q$-conditional  symmetries of the first type listed in
Theorem~\ref{th-1} can be applied to construct exact solutions of
the corresponding  HGF systems of the form (\ref{1-1}). Here we have
examined two special    cases  (\ref{1-1}), when the corresponding
system admits either  the symmetry operators $Q^u_3$ and  $Q^w_3$,
 or the operator $Q^u_4$.

Note that  the  HGF system (\ref{2-3*}) admitting the symmetry
operators $Q^u_3$ and  $Q^w_3$ has the same structure as a system
examined in \cite{ch-dav-2021}. The only difference is such that the
third equation in system (3.2) \cite{ch-dav-2021} contains the terms
$-\frac{d_3}{d_1}uw-\frac{d_3}{d_1}vw$ (with $d_3\not=d_1$) instead
of $-uw-vw$. However, the exact solutions obtained for   HGF system
(\ref{2-3*}) have essentially different structure than those derived
for  system  (3.2) \cite{ch-dav-2021}.

Moreover, some of  the solutions derived in Section~\ref{sec-3}
possess attractive properties allowing us to provide a plausible
archeological
 interpretation (following the terminology used in  \cite{fu-mimura-2021}, it is reasonable
to replace `biological' by  `archeological'). As it is shown by
numerical simulations in  \cite{fu-mimura-2021}, a  typical
asymptotic behavior of solutions of the HGF system (\ref{1-1}) with
the no-flux boundary conditions has either the form  (\ref{3-11}),
or \begin{equation}\label{4-1} (u,\,v,\,w) \rightarrow (0,\, v_0,\,
w_0) \ \texttt{as} \ t \rightarrow +\infty,\end{equation} where
$v_0$ and $ w_0$ are expressed via the system coefficients. The
solutions derived in Subsection~\ref{subsec-3.1}  possess (under the
relevant restrictions) only  the asymptotic behavior  (\ref{3-11})
(obviously (\ref{3-11*}) and (\ref{3-16}) are the same formulae up
to notations). It is in agreement with the numerical results
obtained in \cite{fu-mimura-2021} because we examined the HGF system
(\ref{1-1}) with $a_3=0$ (see system (\ref{2-3})). It means that the
case $a_3<a_4=a_5$ was studied, which predicts the extinction of
hunter-gatherers.

In order to construct exact solutions with the asymptotic behavior
(\ref{4-1})  and to compare with the results obtained in the recent
studies \cite{elia-mimura-2021,fu-mimura-2021}, it is necessary to
examine the HGF system (\ref{1-1}) with $a_3>a_4=a_5$. Such systems
occur among systems (\ref{2-1}) and (\ref{2-2}), which possess
nontrivial $Q$-conditional symmetries. Here we restricted ourselves
on examination of the HGF system (\ref{2-2}) because
 $d_1=d_2$ is assumed in \cite{elia-mimura-2021} and \cite{fu-mimura-2021}. Thus, using
the $Q$-conditional symmetry  operator $Q^u_4$, two families  of
exact solutions  were constructed in Subsection~\ref{subsec-3.2}. In
particular, it was shown how an exact solution of the form
(\ref{3-30})  satisfying the no-flux conditions (\ref{3-31}) at a
bounded interval $\Omega_{ab}$ can be constructed. Moreover, the
solution has the asymptotical behavior (\ref{3-32}) provided
$a_4<a_3=1$ in the HGF system (\ref{2-2}).
 Using archeological terminology, this solution
predicts  the coexistence of  converted farmers  and
hunter-gatherers and  the total extinction of initial  farmers. It
is in agreement with theoretical and numerical results obtained in
\cite{elia-mimura-2021,fu-mimura-2021}.

Exact solutions of the form (\ref{3-34}) do not satisfy the
asymptotical  condition  (\ref{4-1})  (independently on a specific
form of the function $f(t,x)$). On the other hand, these solutions
with a correctly-specified function $f(t,x)$  have the asymptotical
behavior (\ref{3-35}). Thus, such solutions predict the coexistence
of initial  and  converted    farmers and  the total extinction of
hunter-gatherers.
 It is  again in agreement with the results of
\cite{elia-mimura-2021} and \cite{fu-mimura-2021}  because our
solutions are valid for  the HGF system (\ref{2-2}) with the
restriction $a_3=a_4-a_1-1$ , hence $a_4>a_3$. Interestingly, the
multiplier $f_0^\infty$ in  (\ref{3-35}) can be a function of the
space variable $x$, hence a spatial segregation of  initial and
converted farmers  may occur as $t \rightarrow +\infty $.

\medskip
\medskip
\textbf{Acknowledgments} \medskip \\
The authors acknowledge a partial financial support within the
framework of the priority program for research and
scientific-and-technical (experimental) development of the
mathematical department of the NAS of Ukraine in 2022--2023 (Reg. No
0122U000670).

\end{document}